\documentclass[a4paper,fleqn]{cas-sc}

\usepackage[numbers]{natbib}
\usepackage{xcolor}
\usepackage{soul}

\def\tsc#1{\csdef{#1}{\textsc{\lowercase{#1}}\xspace}}
\tsc{WGM}
\tsc{QE}
\tsc{EP}
\tsc{PMS}
\tsc{BEC}
\tsc{DE}

\begin{document}
\let\WriteBookmarks\relax
\def\floatpagepagefraction{1}
\def\textpagefraction{.001}

\shorttitle{Generalised Harmonic Domain Analysis for Transformer Core Hysteresis Modelling}

\shortauthors{J. Schipper et~al.}

\title [mode = title]{Generalised Harmonic Domain Analysis for Transformer Core Hysteresis Modelling}                      



%

\author[1]{Josh Schipper}[
    orcid=0000-0003-3371-0917]

\cormark[1]


\ead{josh.schipper@epecentre.ac.nz}


\credit{Conceptualization, Data curation, Formal analysis, Methodology, Software, Validation, Visualization, Writing - original draft}

\affiliation[1]{organization={Electric Power Engineering Centre, University of      Canterbury},
    addressline={20 Kirkwood Avenue}, 
    city={Upper Riccarton, Christchurch},
    postcode={8041}, 
    country={New Zealand}}

\author[1]{ Radnya Mukhedkar}

\credit{Funding acquisition, Supervision, Writing - review \& editing}

\author[2]{ Neville Watson}[
   orcid=0000-0002-1320-3101]
   

\credit{Funding acquisition, Supervision, Writing - review \& editing}

\affiliation[2]{organization={Department of Electrical and Computer Engineering, University of Canterbury},
    addressline={20 Kirkwood Avenue}, 
    city={Upper Riccarton, Christchurch},
    postcode={8041}, 
    country={New Zealand}}

\author[1]{ Veerabrahmam Bathini}[]

\credit{Writing - review \& editing}

\author[3]{ Jan Meyer}[
   orcid=0000-0002-6884-5101]

\affiliation[3]{organization={Institute of Electrical Power Systems and High Voltage Engineering, Dresden University of Technology},
    addressline={Mommsenstraße 10},
    city={Dresden},
    postcode={01069}, 
    country={Germany}}

\credit{Resources, Writing - review \& editing}

\cortext[cor1]{Corresponding author}



\begin{abstract}
This work identifies the general approach for linearising any power system component in the harmonic domain, that is with respect to its Fourier series coefficients. This ability enables detailed harmonic analysis, and is key as more power electronic devices inject harmonic currents into the power system to its shared detriment. The general approach requires a time domain model of the component, and is most applicable where a conversion to the frequency domain is impractical prior to linearisation. The outcome is a Norton equivalent current source, which expresses linear coupling between harmonic frequencies with admittance matrices. These are the so-called frequency coupling matrices. The general approach is demonstrated for magnetic hysteresis, where a Preisach model has been developed for this purpose. A new data driven approach is used to fit the test results of a small physical transformer to the Preisach model. Results show an improved accuracy in the frequency coupling matrices over models that only considered magnetic saturation. Maximum improvement is observed in the odd harmonic current to odd harmonic voltage couplings. 



\end{abstract}


\begin{highlights}
\item Developed general methodology for creating harmonic domain Norton equivalents.
\item Adapted classical Preisach model of hysteresis for harmonic analysis.
\item Created approximate method for implementing Preisach model from symmetric test data.
\item Improved frequency coupling matrix for transformer open-circuit characteristics.
\end{highlights}


\maketitle

\section{Introduction}

Simulating the propagation of harmonic currents and voltages is valuable to understanding the impact of nonlinear devices on power quality. The increasing prevalence of converter technologies in a hybrid AC/DC landscape means that harmonic analysis will continue to be an important tool \cite{Wilsun2024}, such as in assessing the impact of Geomagnetically Induced Currents on converters \cite{Han2024}. The complex and multivariate nature of harmonic interactions makes linear approximation the basis for tractable frequency domain simulation, where the interactions of two or more nonlinear devices is studied. Linearisation creates Norton equivalent current sources, where a devices terminal voltage and current injection are expressed with Fourier series coefficients. This process of model development is called harmonic domain analysis. The admittance matrix of the Norton equivalent is a frequency coupling matrix.

Harmonic domain analysis provides insight into harmonic coupling that cannot be gained from time domain simulation such as its numerical implementation in Electromagnetic Transient (EMT) simulation. Norton equivalents of multiple nonlinear devices can be combined together with an electrical network to be solved linearly. The result traces feedbacks between nonlinear devices and network elements. Furthermore, simulation accuracy can be improved by updating Norton equivalents in an iterative process by repeat linearisation about each new operating voltage. Combined with generation and load power requirements, a harmonic power-flow is developed \cite{Arrillaga1995}. Alternatively, a harmonic power-flow methodology can apply the linearisation process over an entire network to form a Jacobian matrix within a Newton-Raphson approach \cite{Arrillaga2004}. 

Three methods exist for creating the linear approximation needed for a frequency coupling admittance matrix. The first is physically testing a nonlinear device \cite{Gallo2018, Collin2019}. For each harmonic voltage, a small perturbation is applied around a point of nominal operation to estimate the slope of the harmonic coupling to the current injection. The second method is similar to the first, where multiple small perturbations are applied to a device in a time domain simulation \cite{Yadav2022}. The third method differentiates a frequency domain model of the device \cite{Arrillaga1995, Acha2001}.

Comparing the two simulation methods, the time domain approach is computationally slower as each perturbation has to be simulated with a sufficiently small time step and with some settling time to reach steady-state. Its advantage comes from simpler model development with established simulation tools. The advantage of the frequency domain method is better control of numerical error, where symbolic differentiation can be performed for the Taylor series linearisation. Time-domain methods introduce error from approximating the linearisation through perturbations and in the simulation process. 

Prior to this work, time domain methods had the additional advantage that they can simulate nonlinear devices that cannot be described in the frequency domain in closed-form expressions. For example, the Preisach model of hysteresis is not easily converted to the frequency domain. The contribution of this work shows that a frequency coupling matrix can be linearised in the frequency domain from a time domain model. A general process is established on the first principles of Fourier series analysis. The methodology is demonstrated for transformer magnetic hysteresis, where the classical Preisach model is adapted for time-periodic inputs. These methods and models also have value for the study of ferroresonance \cite{Zare2008} and the frequency response of dynamic systems with hysteresis, where Play operators have been analysed previously \cite{Esbrook2012, Lei2016, Lei2018}.

Hysteresis models have been extensively developed for time domain simulation including for commercial tools \cite{Frame1982, Chandrasena2004, Zirka2015, Zare2015}. This work's second contribution, in adapting the classical Preisach model for harmonic analysis, is the development of a new data fitting method for model implementation.

The work begins with the harmonic domain linearisation of nonlinear devices in Section \ref{sec_general_harmonic_domain}. Section \ref{sec_preisach_model_start} introduces the classical Preisach model and its modification for time-periodic functions. The theoretical basis for fitting test data to the Preisach model is presented in Section \ref{sec_fitting_reduced}. The linearisation of the time-periodic Preisach model is achieved in Section \ref{sec_preisach_harmonic_domain}. Section \ref{sec_results_main} presents the results of applying the techniques to a test transformer.





\section{Generalised Harmonic Domain Analysis}
\label{sec_general_harmonic_domain}

This section generalises the process of developing harmonic domain Norton equivalents. Linearisation is explicitly shown for devices described by operators of time-periodic signals, i.e. $i(t)=f\{v(t)\}$ where $f: \Omega \rightarrow \Omega$ and $\Omega$ is the set of continuous functions with the periodicity property $v(t) = v(t + T)$. Curly braces are a chosen to simplify the notation of operators and make a distinction from functions $i(t)=f(v(t))$ that map $f: \mathbb{R} \rightarrow \mathbb{R}$. Once the general process is explained for $i(t)=f\{v(t)\}$, the linearisation is applied to the well known example of $i(t)=f(v(t),t)$ with the harmonic phasor series (HPS). The general process can be applied to any form of Fourier series, such as the complex Fourier series.

The HPS for current and voltage are:

\begin{equation} \label{eq_vdef}
i(t)=\Re \Bigg\{\sum_{n=0}^{\infty}I_{n}e^{jn \omega t} \Bigg\}, \; v(t)=\Re \Bigg\{\sum_{m=0}^{\infty}V_{m}e^{jm \omega t} \Bigg\}
\end{equation}

\noindent where $I_{n}$ and $V_{m}$ are the harmonic coefficients, $\omega = 2\pi/T$, and $\Re\{\bullet\}$ is the real component. The HPS could have been alternatively defined as the imaginary component of each harmonic phasor. There is no material difference between taking the real or imaginary component except for a difference in phase. The HPS coefficients are calculated according to the transformation from $i(t)$ to $I_{n}$:

\begin{equation} \label{eq_i_coeff_V1}
    I_{n}=\frac{2-\delta[n]}{T} \int_{0}^{T}i(t)e^{-jn\omega t}dt
\end{equation}

\noindent where

\begin{equation} \label{eq_delta_func}
\delta[n]=\left\{
\begin{matrix}
1, & n=0 \\
0, & \text{otherwise}
\end{matrix}
\right.
\end{equation}

General harmonic domain analysis directly applies the Fourier series definition and transformation in the linearisation of harmonic coefficients. A formula for the current harmonic coefficients in terms of the voltage harmonic coefficients is generated by substituting the device model $i(t)=f\{v(t)\}$ and $v(t)$ HPS from (\ref{eq_vdef}) into (\ref{eq_i_coeff_V1}):

\begin{equation} \label{eq_i_coeff_V2}
    I_{n}(\mathbf{V})=\frac{2-\delta[n]}{T} \int_{0}^{T}f\Bigg\{\Re \Bigg\{\sum_{m=0}^{\infty}V_{m}e^{jm \omega t} \Bigg\}\Bigg\}e^{-jn\omega t}dt
\end{equation}

\noindent where the harmonic coefficients are grouped into vectors:


\begin{equation} \label{eq_phasor_vector_def}
\mathbf{I} = \begin{bmatrix}
I_{0}, I_{1}, I_{2}, \cdots
\end{bmatrix}^{T} \; \; \text{and} \; \; \mathbf{V} = \begin{bmatrix}
V_{0}, V_{1}, V_{2}, \cdots
\end{bmatrix}^{T}
\end{equation}

The harmonic domain model is created by linearising $\mathbf{I}(\mathbf{V})$ about the base operating point $\mathbf{I}_{\text{B}} = \mathbf{I}(\mathbf{V}_{\text{B}})$, which should be centred around a wide range of possible operation for the nonlinear device. If (\ref{eq_i_coeff_V2}) has a complex partial derivative with respect to $V_{m}$, then the linear approximation based on forming the first order Taylor series is:

\begin{equation} \label{eq_p_indep}
\Delta\mathbf{I} = Y^{(1)}\Delta\mathbf{V} \; \; \; \text{where} \; \; \; Y^{(1)}_{n,m} = \left. \frac{\partial I_{n}}{\partial V_{m}}\right|_{\mathbf{V}=\mathbf{V}_{\text{B}}}
\end{equation}

The Norton equivalent is created by substituting in the perturbation definitions $\Delta\mathbf{I}=\mathbf{I}-\mathbf{I}_{\text{B}}$ and $\Delta\mathbf{V}=\mathbf{V}-\mathbf{V}_{\text{B}}$ into (\ref{eq_p_indep}). A complex partial derivative exists if the following limit exists for a complex function, $f(z)$:

\begin{equation} \label{eq_complex_der}
f'(z_{0}) = \lim_{z \rightarrow z_{0}} \frac{f(z) - f(z_{0})}{z - z_{0}}
\end{equation}

For real functions this limit exists if the left and right side limits exist and are identical, i.e. approaching the same value from the left and right sides of the real number line. For the complex limit (\ref{eq_complex_der}), instead of a real number line, $z_{0}$ can be approached from any direction in the complex plane \cite{Marshall2019}, where the limit must be identical from every direction. The complex partial derivative exists for simple electrical components, such as resistors, capacitors, inductors, and combinations of these components. However, for most power electronic converters and other nonlinear components the complex partial derivative does not exist, and to form a linear approximation it is necessary to compute the partial derivatives with the complex coefficients split into their real and imaginary components:

\begin{equation} \label{IV_split}
   I_{n}=I_{n}^{\Re} + jI_{n}^{\Im} \; \text{and} \; V_{m}=V_{m}^{\Re} + jV_{m}^{\Im}
\end{equation}

Therefore, (\ref{eq_i_coeff_V2}) is split into two real equations:

\begin{equation} \label{eq_i_coeff_V2_R}
    I_{n}^{\Re}(\mathbf{V})=\frac{2-\delta[n]}{T} \int_{0}^{T}f\Bigg\{\Re \Bigg\{\sum_{m=0}^{\infty}V_{m}e^{jm \omega t} \Bigg\}\Bigg\}\cos{(n\omega t)}\, dt
\end{equation}
\begin{equation} \label{eq_i_coeff_V2_I}
    I_{n}^{\Im}(\mathbf{V})=-\frac{2-\delta[n]}{T} \int_{0}^{T}f\Bigg\{\Re \Bigg\{\sum_{m=0}^{\infty}V_{m}e^{jm \omega t} \Bigg\}\Bigg\}\sin{(n\omega t)}\, dt
\end{equation}

The first order Taylor series of (\ref{eq_i_coeff_V2_R}) and (\ref{eq_i_coeff_V2_I}) is:

\begin{equation} \label{eq_I_real_taylor}
    \Delta I_{n}^{\Re} = \sum_{m=0}^{\infty} \Bigg( \frac{\partial I_{n}^{\Re}}{\partial V_{m}^{\Re}}\Delta V_{m}^{\Re} + \frac{\partial I_{n}^{\Re}}{\partial V_{m}^{\Im}}\Delta V_{m}^{\Im} \Bigg)
\end{equation}
\begin{equation} \label{eq_I_imag_taylor}
    \Delta I_{n}^{\Im} = \sum_{m=0}^{\infty} \Bigg( \frac{\partial I_{n}^{\Im}}{\partial V_{m}^{\Re}}\Delta V_{m}^{\Re} + \frac{\partial I_{n}^{\Im}}{\partial V_{m}^{\Im}}\Delta V_{m}^{\Im} \Bigg)
\end{equation}

\noindent where each partial derivative is evaluated at $v(t) = v_{B}(t)$. Whether and how each partial derivative can be calculated depends on the definition of the chosen operator and $v_{B}(t)$. Section \ref{sec_harmonic_nonlinear} evaluates the partial derivatives for the function $i(t)=f(v(t),t)$. Then the process for the Preisach operator is described in Section \ref{sec_preisach_harmonic_domain}. Before this, the meaning of the linearisation is elaborated upon. The two linearisations of (\ref{eq_I_real_taylor}) and (\ref{eq_I_imag_taylor}) can be combined into the complex coefficients of the HPS:

\begin{equation} \label{eq_p_dep}
\Delta\mathbf{I} = Y^{(1)}\Delta\mathbf{V} + Y^{(2)}\overline{\Delta\mathbf{V}}
\end{equation}

\noindent where the vinculum expresses conjugation, and:

\begin{equation} \label{eq_y1def_p_dep}
   Y^{(1)}_{n,m} = \frac{1}{2}\Bigg(\frac{\partial I_{n}^{\Re}}{\partial V_{m}^{\Re}} + \frac{\partial I_{n}^{\Im}}{\partial V_{m}^{\Im}} \Bigg) + \frac{j}{2}\Bigg(\frac{\partial I_{n}^{\Im}}{\partial V_{m}^{\Re}} - \frac{\partial I_{n}^{\Re}}{\partial V_{m}^{\Im}} \Bigg)
\end{equation}
\begin{equation} \label{eq_y2def_p_dep}
   Y^{(2)}_{n,m} = \frac{1}{2}\Bigg(\frac{\partial I_{n}^{\Re}}{\partial V_{m}^{\Re}} - \frac{\partial I_{n}^{\Im}}{\partial V_{m}^{\Im}} \Bigg) + \frac{j}{2}\Bigg(\frac{\partial I_{n}^{\Im}}{\partial V_{m}^{\Re}} + \frac{\partial I_{n}^{\Re}}{\partial V_{m}^{\Im}} \Bigg)
\end{equation}

This form shows that the partial derivatives are admittances. Furthermore, if the complex partial derivative in (\ref{eq_p_indep}) exists, then $\partial I_{n}/\partial V_{m}$ is equivalent to (\ref{eq_y1def_p_dep}) and $Y^{(2)}_{n,m} = 0$. Note, $Y^{(2)}_{n,m} = 0$ is the Cauchy-Riemann equations.

The HPS's main advantage over other forms of Fourier series is a direct connection between the concept of phase dependency and the existence of the complex partial derivative. Equation (\ref{eq_p_indep}) is a phase independent linear approximation because the admittance $\Delta I_{n}/\Delta V_{m}$ is constant and not a function of the magnitude or phase of $\Delta V_{m}$, while assuming all other $\Delta V_{k}$ are zero with $k\neq m$. Whereas for (\ref{eq_p_dep}), $\Delta I_{n}/\Delta V_{m}=Y^{(1)}_{n,m}+Y^{(2)}_{n,m}(\overline{\Delta V_{m}}/\Delta V_{m})$ is a function of the phase of $\Delta V_{m}$. For the phase dependent case, if the harmonic admittance $\Delta I_{n}/\Delta V_{m}$ is plotted for variations in the phase angle of $\Delta V_{m}$ from 0 to 360\textdegree, then $\Delta I_{n}/\Delta V_{m}$ twice trace a circle in the complex admittance plane, where the centre of the circle is $Y^{(1)}_{n,m}$ and the radius is equal to $|Y^{(2)}_{n,m}|$. For an expanded discussion on this topic, \cite{Malagon2024} provides a link to Wirtinger Calculus.

\subsection{Harmonic Domain for Nonlinear Function}
\label{sec_harmonic_nonlinear}

This section applies generalised harmonic domain analysis to the nonlinear function $f\{v(t)\}=y(v(t),t)$ with the periodicity property $y(v,t)=y(v,t+T)$. This step is required for forming the frequency coupling matrix for the Preisach model, but is here presented as an example before applying it to an operator in Section \ref{sec_preisach_harmonic_domain}. The first formulations of the harmonic domain \cite{Arrillaga1995} have analysed nonlinear functions with the complex Fourier series, which resulted in a Toeplitz matrix for the frequency coupling matrix. This section verifies the equivalent result for the HPS. 

Previous formulations began by differentiating $y(v,t)$ with respect to $v$ to obtain the time domain expression of the linearisation: 

\begin{equation} \label{eq_spec_case_hd}
\Delta i(t) = y_{v}(v_{\text{B}}(t), t) \, \Delta v(t) \; \; \text{where} \; \; y_{v} = \frac{\partial y}{\partial v}
\end{equation}

\noindent and $\Delta i(t)$, $\Delta v(t)$, and $v_{\text{B}}(t)$ are time periodic signals generated from the harmonic coefficients of $\Delta\mathbf{I}$, $\Delta\mathbf{V}$, and $\mathbf{V}_{\text{B}}$, respectively. These formulations then apply the convolution formula to (\ref{eq_spec_case_hd}) to convert to the harmonic domain. Although this approach gives the correct result, it misses the step demonstrating that $\Delta i(t)$ from (\ref{eq_spec_case_hd}) and $\Delta\mathbf{I}$ from (\ref{eq_p_dep}) are equivalent through the transformation (\ref{eq_i_coeff_V1}). Therefore, generalised hamonic domain analysis will improve upon previous attempts to clarify this tension \cite{Smith1996a}, where the equivalence of (\ref{eq_p_dep}) and (\ref{eq_spec_case_hd}) will be demonstrated.

The nonlinear function $i(t)=y(v(t),t)$ is substituted into (\ref{eq_i_coeff_V1}):

\begin{equation} \label{eq_i_coeff_V4}
    I_{n}=\frac{2-\delta[n]}{T} \int_{0}^{T}y(v(t),t)e^{-jn\omega t}dt
\end{equation}

The complex partial derivatives, $\partial I_{n}/\partial V_{m}$, generally do not exist for (\ref{eq_i_coeff_V4}); therefore, the voltage signal $v(t)$ is separated into real and imaginary components:

\begin{equation} \label{eq_v_sep_realimag}
v(t) = \sum_{m=0}^{\infty}\bigg(V_{m}^{\Re}\cos{(m\omega t)} - V_{m}^{\Im}\sin{(m\omega t)} \bigg)
\end{equation}

To simplify calculating the partial derivatives of (\ref{eq_i_coeff_V4}) with respect to the real and imaginary harmonic voltages, the symbol $\psi$ represents each $V_{m}^{\Re}$ or $V_{m}^{\Im}$ one at a time. The Leibniz integral rule brings the differentiation within the integral term of (\ref{eq_i_coeff_V4}):

\begin{equation} \label{eq_hps_ipartial_gen}
    \frac{\partial I_{n}}{\partial \psi} = \frac{2-\delta[n]}{T} \int_{0}^{T}y_{v}(v_{\text{B}}(t),t)\frac{\partial v}{\partial \psi}e^{-jn\omega t}dt
\end{equation}

\noindent where the partial derivative is evaluated at $v(t) = v_{B}(t)$. For the case of $\psi = V_{0}^{\Re}$, differentiating (\ref{eq_v_sep_realimag}) means $\partial v/\partial \psi = 1$ and (\ref{eq_hps_ipartial_gen}) becomes:

\begin{equation} \label{eq_hps_ipartial_case1}
\frac{\partial I_{n}}{\partial V_{0}^{\Re}} = \frac{2-\delta[n]}{T} \int_{0}^{T}y_{v}(v_{\text{B}}(t),t)e^{-jn\omega t}dt=Y_{n}
\end{equation}

The admittances $Y_{n}$ are the HPS coefficients for $y_{v}(v_{\text{B}}(t),t)$ according to the transformation definition (\ref{eq_i_coeff_V1}). For the case of $\psi = V_{0}^{\Im}$, $\partial v/\partial \psi = 0$ and results in $\partial I_{n}/\partial V_{0}^{\Im} = 0$. For the case of $\psi = V_{m}^{\Re}$ with $m>0$, the derivative is $\partial v/\partial \psi = \cos{(m\omega t)}$ and (\ref{eq_hps_ipartial_gen}) becomes:

\begin{equation} \label{eq_hps_ipartial_case2}
\frac{\partial I_{n}}{\partial V_{m}^{\Re}} = \frac{2-\delta[n]}{T} \int_{0}^{T}y_{v}(v_{\text{B}}(t),t)\cos{(m\omega t)}e^{-jn\omega t}dt
\end{equation}

To have an expression for (\ref{eq_hps_ipartial_case2}) in terms of $Y_{n}$, the identity $\cos{(\theta)} = (e^{j\theta} + e^{-j\theta})/2$ is applied:

\begin{equation} \label{eq_hps_ipartial_case3}
\frac{\partial I_{n}}{\partial V_{m}^{\Re}} = \frac{1}{2}\frac{2-\delta[n]}{T} \int_{0}^{T}y_{v}(v_{\text{B}}(t),t)\Big(e^{-j(n-m)\omega t} + e^{-j(n+m)\omega t}\Big) dt
\end{equation}

Further simplification makes a distinction between when $n\geq m$ and $n < m$ as only positive harmonic coefficients are defined in the HPS for $y_{v}(v_{\text{B}}(t),t)$. For $n\geq m$:

\begin{equation} \label{eq_hps_ipartial_case4}
\frac{\partial I_{n}}{\partial V_{m}^{\Re}} = \frac{1}{2}\Bigg(\frac{2-\delta[n]}{2-\delta[n-m]}Y_{n-m} + \frac{2-\delta[n]}{2-\delta[n+m]}Y_{n+m}\Bigg)
\end{equation}

\noindent and for $n < m$:

\begin{equation} \label{eq_hps_ipartial_case5}
\frac{\partial I_{n}}{\partial V_{m}^{\Re}} = \frac{1}{2}\Bigg(\frac{2-\delta[n]}{2-\delta[n-m]}\overline{Y_{m-n}} + \frac{2-\delta[n]}{2-\delta[n+m]}Y_{n+m}\Bigg)
\end{equation}

Lastly, following a similar process for $\psi = V_{m}^{\Im}$ with $m>0$, $\partial v/\partial \psi = -\sin{(m\omega t)}$ and (\ref{eq_hps_ipartial_gen}) becomes:

\begin{equation} \label{eq_hps_ipartial_case8}
\frac{\partial I_{n}}{\partial V_{m}^{\Im}} = -\frac{1}{2j}\Bigg(\frac{2-\delta[n]}{2-\delta[n-m]}Y_{n-m} - \frac{2-\delta[n]}{2-\delta[n+m]}Y_{n+m}\Bigg)
\end{equation}

\noindent for $n \geq m$ and for $n < m$:

\begin{equation} \label{eq_hps_ipartial_case9}
\frac{\partial I_{n}}{\partial V_{m}^{\Im}} = -\frac{1}{2j}\Bigg(\frac{2-\delta[n]}{2-\delta[n-m]}\overline{Y_{m-n}} - \frac{1}{2j}\frac{2-\delta[n]}{2-\delta[n+m]}Y_{n+m}\Bigg)
\end{equation}

The final step is the construction of $Y^{(1)}$ and $Y^{(2)}$, which for unseparated $I_{n}$ can be obtained from simplifying (\ref{eq_y1def_p_dep}) and (\ref{eq_y2def_p_dep}):

\begin{equation} \label{eq_y12def_p_dep_simp}
   Y^{(1)}_{n,m} = \frac{1}{2} \frac{\partial I_{n}}{\partial V_{m}^{\Re}} - \frac{j}{2} \frac{\partial I_{n}}{\partial V_{m}^{\Im}} \quad \text{and} \quad Y^{(2)}_{n,m} = \frac{1}{2} \frac{\partial I_{n}}{\partial V_{m}^{\Re}} + \frac{j}{2} \frac{\partial I_{n}}{\partial V_{m}^{\Im}}
\end{equation}

Therefore, the linearisation for the nonlinear function $i(t)=y(v(t),t)$ is: 

\begin{equation} \label{eq_Y1def_for_conv}
Y^{(1)}=\frac{1}{2}\begin{bmatrix} 
Y_{0} & \overline{Y_{1}}/2 & \overline{Y_{2}}/2 & \overline{Y_{3}}/2 & \cdots \\
Y_{1} & 2Y_{0} & \overline{Y_{1}} & \overline{Y_{2}} & \cdots \\
Y_{2} & Y_{1} & 2Y_{0} & \overline{Y_{1}} & \cdots \\
Y_{3} & Y_{2} & Y_{1} & 2Y_{0} & \cdots \\
\vdots & \vdots & \vdots & \vdots & \ddots \\
\end{bmatrix}
\end{equation}

\begin{equation} \label{eq_Y2def_for_conv}
Y^{(2)}=\frac{1}{2}\begin{bmatrix} 
Y_{0} & Y_{1}/2 & Y_{2}/2 & Y_{3}/2 & \cdots \\
Y_{1} & Y_{2} & Y_{3} & Y_{4} & \cdots \\
Y_{2} & Y_{3} & Y_{4} & Y_{5} & \cdots \\
Y_{3} & Y_{4} & Y_{5} & Y_{6} & \cdots \\
\vdots & \vdots & \vdots & \vdots & \ddots \\
\end{bmatrix}
\end{equation}

Comparing (\ref{eq_p_dep}) with the convolution formula for the HPS in (\ref{example_conv}) shows that they are the same equation, where for the nonlinear function $Y^{(1)}$ and $Y^{(2)}$ come from (\ref{eq_Y1def_for_conv}) and (\ref{eq_Y2def_for_conv}). This proves the time domain representation of the linearisation (\ref{eq_spec_case_hd}) is correct. The convolution formula is adapted from \cite{Smith1996a} (in Appendix A.2) to use the real component definition of (\ref{eq_vdef}).

\begin{equation} \label{example_conv}
\Delta I_{n} =\frac{1}{2}\sum_{m=0}^{n}\Delta V_{n-m}Y_{m} + \frac{1}{2+2\delta[n]}\sum_{m=0}^{\infty}\big(\Delta V_{n+m}\overline{Y_{m}} + \overline{\Delta V_{m}}Y_{n+m}\big)
\end{equation}

\section{Preisach Model of Hysteresis}
\label{sec_preisach_model_start}

This section describes the basic form of a Preisach model before adapting it to time-periodic inputs. Its implementation from perfect test data is also explained. The Preisach operator has the current $i(t)$ of a coil wound around a core as its input. The output is the equivalent flux linkage $\lambda(t)$ for the same coil in units of Volt-seconds (Vs). This model is a single port model, which will be used to characterise a transformer's open-circuit characteristics. 

\begin{figure}
	\centering
		\includegraphics[scale=1]{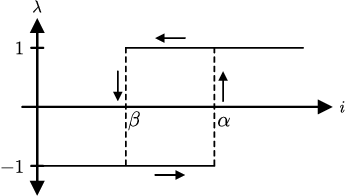}
	\caption{Ideal Relay.}
	\label{fig_ideal_relay}
\end{figure}

The Preisach model of hysteresis is an extension of the ideal relay concept. Fig. \ref{fig_ideal_relay} shows that an ideal relay shifts between two states. Two parameters, $\beta$ and $\alpha$, determine the necessary current, $i$, for the transitioning between these two states. The value $\beta$ is the necessary current to achieve a downward transition, and conversely, if the current is greater than $\alpha$ the upward transition occurs. For transitions between the two states to always be possible it is necessary for $\beta \leq \alpha$. Implicit within the definition of an ideal relay is a history term determining initial state, which is omitted from the notation of this work. The ideal relay is expressed by the operator $r\{i(t); \beta, \alpha\}$.

The classic Preisach model, $H_{s}$, combines multiple ideal relays together in the integral equation:

\begin{equation} \label{eq_preisach_simple}
    \lambda(t) = H_{s}\{i(t)\} = \iint_{\beta \leq \alpha} p(\beta, \alpha) \; r\{i(t); \beta, \alpha\} d\!\beta \, d\!\alpha
\end{equation}

The Preisach model combines infinitesimally small ideal relays of varying settings according to the Preisach weighting function $p(\beta, \alpha)$, which characteristics core properties. The region of integration is the half plane $\beta \leq \alpha$ because of the restriction on $r\{i(t); \beta, \alpha\}$. As the ideal relay has only two outputs of positive and negative one, (\ref{eq_preisach_simple}) can be separated into two integrals with time varying regions of integration:

\begin{equation} \label{eq_preisach_two_regions}
    H_{s}\{i(t)\} = \iint_{S^{+}(t)} p(\beta, \alpha) d\!\beta \, d\!\alpha - \iint_{S^{-}(t)} p(\beta, \alpha) d\!\beta \, d\!\alpha
\end{equation}

\begin{figure}
	\centering
		\includegraphics[scale=1]{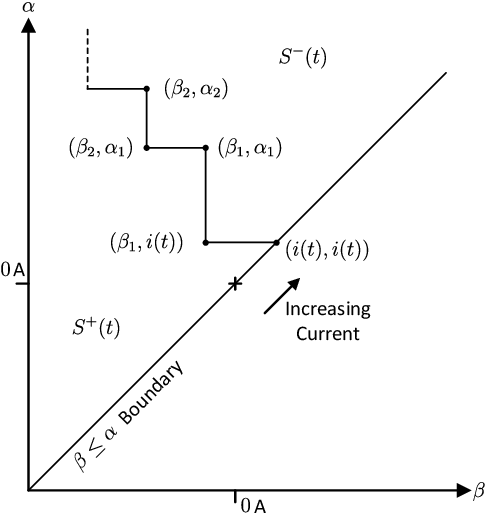}
	\caption{Example domain of $p(\beta, \alpha)$.}
	\label{fig_preisach_simple}
\end{figure}

The positive region $S^{+}(t)$ contains all relays whose settings $(\beta, \alpha) \in S^{+}(t)$ result in the relay being in the positive state. $S^{-}(t)$ is similarly defined for the negative state. An example of these regions is shown in Fig. \ref{fig_preisach_simple}.

The boundary between $S^{+}(t)$ and $S^{-}(t)$ consists of vertical and horizontal lines determined by pairs of prior minimums $\beta_{j}$ and maximums $\alpha_{j}$ from $i(t)$, where $j$ is an index starting from the most recent extrema. The boundary line begins along the $\beta=\alpha$ line at the point $(i(t)), i(t))$ as shown in Fig. \ref{fig_preisach_simple}. If the current is increasing, then the boundary extends horizontally out to the left because all relays that have $\alpha = i(t)$ transition from the negative to the positive state. As the current continues to increase, the horizontal line moves up increasing the size of the $S^{+}(t)$ region and decreasing the size of the $S^{-}(t)$ region. When $i(t)$ reaches a maximum current and begins to decrease, the boundary then extends vertically from the $(i(t)), i(t))$ point, as all relays that have $\beta = i(t)$ transition from the positive to the negative state.

History in the Preisach model is contained in prior maximum and minimum currents, which reflects the history properties of magnetic materials. History can be wiped if $i(t)$ decreases below $\beta_{j}$ or increases past $\alpha_{j}$. The following section will modify the boundaries of $S^{+}(t)$ and $S^{-}(t)$ so that history does not have to be known beyond one period $T$ in the past.

\subsection{Time-Periodic Preisach Model}
\label{sec_time_periodic_preisach}

For classical Preisach models with only time-periodic inputs $i(t) = i(t + T)$, history is reset at least two times every period. This occurs when $i(t)$ equals its minimum $\beta_{m}=\min\{i(t)\}$ or maximum $\alpha_{m}=\max\{i(t)\}$. For each of these times the only history known is $\beta_{m}$ and $\alpha_{m}$, where one is reached at the point of reset, and the other occurs within time $T$ of the past. Since $i(t)$ never goes below $\beta_{m}$ or above $\alpha_{m}$, ideal relays with $\beta < \beta_{m}$ and $\alpha > \alpha_{m}$ will never change state, and can be omitted from the calculation of $\lambda(t)$. The time-periodic Preisach model, $H_{t}$, limits the region of integration to reflect this simplification:

\begin{equation} \label{eq_preisach_time_periodic}
    \lambda(t) = H_{t}\{i(t)\} = c(i(t), \beta_{m}, \alpha_{m}) + \iint_{\beta \leq \alpha}^{\substack{\beta \geq \beta_{m} \\ \alpha \leq \alpha_{m}}} p(\beta, \alpha) \; r\{i(t); \beta, \alpha\} d\!\beta \, d\!\alpha
\end{equation}

\noindent and a common mode function $c(i(t), \beta_{m}, \alpha_{m})$ is also added. Limiting the regional of integration does not materially improve model detail above the classical form. However, it will simplify the calculation of $p(\beta, \alpha)$ from time-periodic test data in Section \ref{sec_perfect_fit}.

The common mode function does extend the capabilities of the time-periodic model above the classical form. Also, including $\beta_{m}$ and $\alpha_{m}$ as inputs provides a further improvement to the typical extension $c(i(t))$. Its advantage is explained in Section \ref{sec_perfect_fit}. Before this, the numerical implementation of the time-period Preisach model is explained with the shape function.

\subsection{Numerical Implementation - Shape Function Form}

Since \cite{Everett1955}, it is known that the Preisach model can be implemented through a summation of Everett functions, which are also called shape functions \cite{Brokate1996}. Here, the form of (\ref{eq_preisach_time_periodic}) is transformed according to \cite{Mayergoyz2003}:

\begin{equation} \label{eq_preisach_characteristic}
    \lambda(t) = H_{t}\{i(t)\} = c(i(t), \beta_{m}, \alpha_{m}) + \epsilon\{\rho_{0}\}h(\beta_{m}, \alpha_{m}) + 2 \!\!\! \sum_{j = 1}^{N(t)-1} \epsilon\{\rho_{j}\}h(\beta\{\rho_{j}\}, \alpha\{\rho_{j}\})
\end{equation}

\noindent where $h(\beta_{m}, \alpha_{m})$ is the shape function with the requirement that $h(\gamma, \gamma) = 0$. The shape function is evaluated at vertices, $\rho_{j} = (\beta_{k}, \alpha_{l})$, on the boundary line between $S^{+}(t)$ and $S^{-}(t)$, as shown in Fig. \ref{fig_preisach_simple}. The first point is $\rho_{0} = (\beta_{m}, \alpha_{m})$, and the last point is $\rho_{N(t)} = (i(t), i(t))$. The number of points $N(t)$ can fluctuate with time as history is wiped or when current reverses direction. All remaining points fall into two categories: those that are on the tip of the staircase called upward points, and those that are in the crevice of the staircase called downward points. Upward points have a positive contribution while downward points have a negative contribution. To identify the type of each point, the operator $\epsilon\{\rho_{j}\}$ is created:

\begin{equation}
    \epsilon\{\rho_{j}\} = \left\{ \begin{array}{ll}
    1 & \quad \text{if } \alpha\{\rho_{j}\} \text{ occurs before } \beta\{\rho_{j}\} \text{ in time } t\\
    -1 & \quad \text{if } \beta\{\rho_{j}\} \text{ occurs before } \alpha\{\rho_{j}\} \text{ in time } t
    \end{array}\right.
\end{equation}

\noindent where $\beta\{\rho_{j}\}$ returns the first value of $\rho_{j}$, which is a historical minimum $\beta_{k}$. The operator $\alpha\{\rho_{j}\}$ returns the second value of $\rho_{j}$, which is a historical maximum $\alpha_{l}$. The second to last point, $\rho_{N(t)-1}$, requires special consideration. If $i(t)$ is increasing then $\rho_{N(t)-1}=(\beta_{1}, i(t))$. If $i(t)$ is decreasing then $\rho_{N(t)-1}=(i(t), \alpha_{1})$. Note, (\ref{eq_preisach_characteristic}) does not include the last point $\rho_{N(t)}$ because $h(i(t), i(t)) = 0$.

The shape function is closely related to the Preisach weighting function \cite{Mayergoyz2003}, which is split into two components $p(\beta, \alpha) = \mu(\beta, \alpha) + \eta(\alpha)\delta(\alpha - \beta)$:

\begin{equation} \label{eq_general_preisach_weight_formulas}
    \mu(\beta, \alpha) = \left. \frac{\partial^{2}h}{\partial \beta_{m} \, \partial \alpha_{m}} \right|_{\substack{\beta_{m} = \beta \\ \alpha_{m} = \alpha}} \quad \text{and} \quad \eta(\alpha) = \left. \frac{\partial h}{\partial \beta_{m}} \right|_{\substack{\beta_{m} = \alpha \\ \alpha_{m} = \alpha}} = -\left. \frac{\partial h}{\partial \alpha_{m}} \right|_{\substack{\beta_{m} = \alpha \\ \alpha_{m} = \alpha}}
\end{equation}

The first component $\mu(\beta, \alpha)$ is the distributed Preisach weighting function, while the weight of the second component is entirely placed along the boundary line $\beta=\alpha$. The unit impulse function $\delta(\alpha - \beta)$ is defined so that:

\begin{equation} \label{eq_shaping_conversion}
    \int_{\beta_{m}}^{\alpha_{m}} \int_{\beta_{m}}^{\alpha} \eta(\alpha)\delta(\alpha - \beta) d\!\beta \, d\!\alpha = \int_{\beta_{m}}^{\alpha_{m}}\eta(\alpha)d\!\alpha
\end{equation}

The advantage of the shape function form is evident. The numerical integration of (\ref{eq_preisach_time_periodic}) is slower than the evaluation of the shaping function in (\ref{eq_preisach_characteristic}). Also, the formation of $p(\beta, \alpha)$ requires differentiation as shown in (\ref{eq_shaping_conversion}), which is numerically prone to error. Section \ref{sec_perfect_fit} shows that $c(i(t), \beta_{m}, \alpha_{m})$ and $h(\beta_{m}, \alpha_{m})$ can be constructed without differentiation.

\subsection{Fitting Perfect Test Data}
\label{sec_perfect_fit}

\begin{figure}
	\centering
		\includegraphics[scale=1]{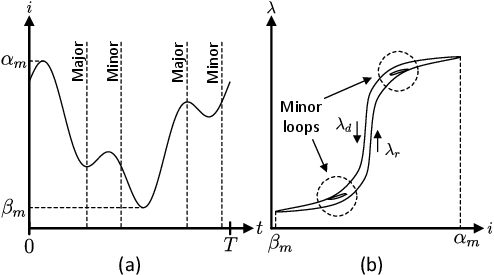}
	\caption{(a) Example magnetizing current waveform showing the different times major and minor hysteresis loops are created. (b) Example hysteresis loop for the current from (a) with minor loops shown within the major loop.}
	\label{fig_major_minor_diff}
\end{figure}

The shape function $h(\beta_{m}, \alpha_{m})$ and the common mode function $c(i(t), \beta_{m}, \alpha_{m})$ are constructed from a series of tests. Monotone sinusoidal currents are injected into a test coil, where the minimum current $\beta_{m}$ and maximum current $\alpha_{m}$ are varied. The resulting equivalent flux linkage, calculated after integrating the terminal voltage, can be plotted on the $(i, \lambda)$ plane to show the major loop hysteresis curves. These are equivalent to the first order reversal curves (FORC) for the classical Preisach model. The major loop results can be separated into two halves: the rising current component $\lambda_{r}(i, \beta_{m}, \alpha_{m})$ and the dropping current $\lambda_{d}(i, \beta_{m}, \alpha_{m})$, as indicated in Fig. \ref{fig_major_minor_diff}(b) for an anti-clockwise orientation.

\begin{figure}
	\centering
		\includegraphics[scale=1]{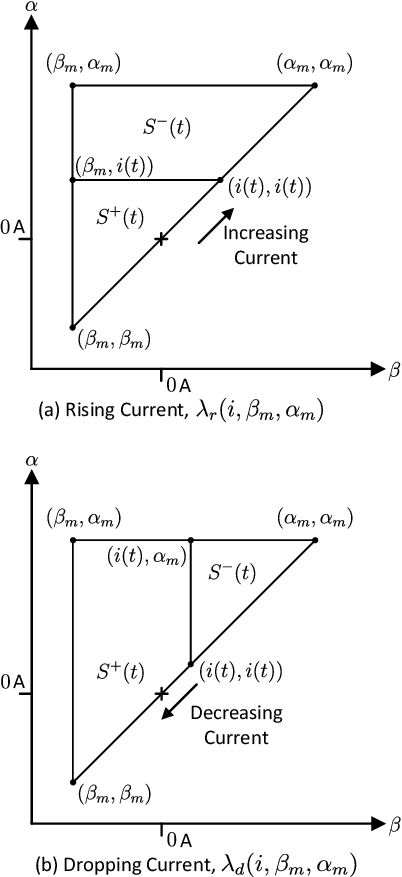}
	\caption{Regions of $p(\beta, \alpha)$ for calculating the centred test data from the time periodic Preisach model.}
	\label{fig_major_loop}
\end{figure}

The equations for the test results are described by (\ref{eq_preisach_characteristic}) with help from Fig. \ref{fig_major_loop}:

\begin{equation}\label{eq_lambda_r_construct}
    \lambda_{r}(i, \beta_{m}, \alpha_{m}) = c(i(t), \beta_{m}, \alpha_{m}) + h(\beta_{m}, \alpha_{m}) - 2h(\beta_{m}, i)
\end{equation}
\begin{equation}\label{eq_lambda_d_construct}
    \lambda_{d}(i, \beta_{m}, \alpha_{m}) = c(i(t), \beta_{m}, \alpha_{m}) - h(\beta_{m}, \alpha_{m}) + 2h(i, \alpha_{m})
\end{equation}

Test data are separated into differential mode components with subscript $s$ and common mode components with subscript $c$:

\begin{equation} \label{eq_diff_conversion}
    \lambda_{s}(i, \beta_{m}, \alpha_{m}) = \frac{1}{2}\big(\lambda_{d}(i, \beta_{m}, \alpha_{m}) - \lambda_{r}(i, \beta_{m}, \alpha_{m}) \big)
\end{equation}
\begin{equation} \label{eq_common_conversion}
    \lambda_{c}(i, \beta_{m}, \alpha_{m}) = \frac{1}{2}\big(\lambda_{d}(i, \beta_{m}, \alpha_{m}) + \lambda_{r}(i, \beta_{m}, \alpha_{m}) \big)
\end{equation}

Substituting (\ref{eq_lambda_r_construct}) and (\ref{eq_lambda_d_construct}) into (\ref{eq_diff_conversion}) derives an implicit definition for $h(\beta_{m}, \alpha_{m})$:

\begin{equation} \label{eq_h_implicit}
    \lambda_{s}(i, \beta_{m}, \alpha_{m}) = h(\beta_{m}, i) + h(i, \alpha_{m}) - h(\beta_{m}, \alpha_{m})
\end{equation}

Section \ref{sec_explicit_h} explains how to convert this implicit definition to an explicit construction. However, conditions determining permissible test results for the differential component are identified in the next section. The remainder of this section derives $c(i(t), \beta_{m}, \alpha_{m})$ assuming $h(\beta_{m}, \alpha_{m})$ is known. Substitute (\ref{eq_lambda_r_construct}) and (\ref{eq_lambda_d_construct}) into (\ref{eq_common_conversion}), and rearrange for the common mode function:

\begin{equation} \label{eq_common_mode_function}
    c(i, \beta_{m}, \alpha_{m}) = \lambda_{c}(i, \beta_{m}, \alpha_{m}) + h(\beta_{m}, i) - h(i, \alpha_{m}) 
\end{equation}

The advantage of including $\beta_{m}$ and $\alpha_{m}$ as arguments to the common mode function is to avoid restricting the acceptable test data. If the common mode function is only a function of current $c(i(t))$, then the right hand side of (\ref{eq_common_mode_function}) would have to be constant with respect to $\beta_{m}$ and $\alpha_{m}$, which could restrict either $\lambda_{c}(i, \beta_{m}, \alpha_{m})$ or $\lambda_{s}(i, \beta_{m}, \alpha_{m})$ through $h(\beta_{m}, \alpha_{m})$. Furthermore, $c(i, \beta_{m}, \alpha_{m})$ allows $h(\beta_{m}, \alpha_{m})$ to be entirely determined by $\lambda_{s}(i, \beta_{m}, \alpha_{m})$ in Section \ref{sec_explicit_h}.

\subsection{Permissible Test Data}
\label{sec_permissible_data}

There are three conditions that restrict permissible test data. The first two conditions can be reasonably assumed. The third condition is a result of the congruency property of the classical Preisach model. Condition 1 requires $\lambda_{r}(i, \beta_{m}, \alpha_{m})$ and $\lambda_{d}(i, \beta_{m}, \alpha_{m})$ to form a loop:   

\begin{equation} \label{eq_loop_bottom}
\lambda_{r}(\beta_{m}, \beta_{m}, \alpha_{m}) =  \lambda_{d}(\beta_{m}, \beta_{m}, \alpha_{m})
\end{equation}
\begin{equation} \label{eq_loop_top}
\lambda_{r}(\alpha_{m}, \beta_{m}, \alpha_{m}) =  \lambda_{d}(\alpha_{m}, \beta_{m}, \alpha_{m})
\end{equation}

This condition can be reasonably assumed because the input current is time-periodic. However, all transients in the measured $\lambda(t)$ must have settled before separating into rising and dropping periods. The implication of Condition 1 is $\lambda_{s}(\beta_{m}, \beta_{m}, \alpha_{m}) = 0$ and $\lambda_{s}(\alpha_{m}, \beta_{m}, \alpha_{m}) = 0$ by (\ref{eq_diff_conversion}). Furthermore, this guarantees $h(\gamma, \gamma) = 0$ through appropriate substitutions to (\ref{eq_h_implicit}).

Condition 2 requires that all first order partial derivatives of $\lambda_{r}(i, \beta_{m}, \alpha_{m})$ and $\lambda_{d}(i, \beta_{m}, \alpha_{m})$ are to be continuous with respect to all arguments. Also, all second order partial derivatives have to be piece-wise continuous. In summary, this condition guarantees that the all first order partial derivatives of $h(\beta_{m}, \alpha_{m})$ are continuous, and all second order partial derivatives are piecewise continuous. Therefore, $h(\beta_{m}, \alpha_{m})$ can be reproduced from the integration of $\mu(\beta, \alpha)$ and $\eta(\alpha)$ in (\ref{eq_preisach_time_periodic}) to obtain (\ref{eq_preisach_characteristic}).

For the classical Preisach model, the congruency property states the shape of hysteresis curve is determined by the value of the latest input extrema. The history of all prior extrema only shifts the curve vertically in the $(i, \lambda)$ plane. For the time-periodic Preisach model, this property is slightly modified by $c(i, \beta_{m}, \alpha_{m})$. To demonstrate, simplify (\ref{eq_preisach_characteristic}) for two different histories, but with the same latest maximum current $\alpha_{1}$ and with current decreasing:

\begin{equation} \label{eq_lambda_history_1}
    \lambda_{1}(i) = c(i, \beta_{m,1}, \alpha_{m,1}) + h_{1} + 2h(i, \alpha_{1}) 
\end{equation}
\begin{equation} \label{eq_lambda_history_2}
    \lambda_{2}(i) = c(i, \beta_{m,2}, \alpha_{m,2}) + h_{2} + 2h(i, \alpha_{1}) 
\end{equation}

\noindent where $h_{1}$ and $h_{2}$ are histories of the Preisach operator, which are constant with respect to $i$. Subtracting (\ref{eq_lambda_history_2}) from (\ref{eq_lambda_history_1}):

\begin{equation} \label{eq_lambda_history_diff}
    \lambda_{1}(i) - \lambda_{2}(i) = c(i, \beta_{m,1}, \alpha_{m,1}) - c(i, \beta_{m,2}, \alpha_{m,2}) + h_{1} - h_{2}
\end{equation}

This implies the shape of the curve can be modified by differences in $\beta_{m}$ and $\alpha_{m}$ by the common mode function. Nonetheless, the hysteresis component of (\ref{eq_preisach_characteristic}) does not modify the shape, which places a restriction on acceptable $\lambda_{s}(i, \beta_{m}, \alpha_{m})$, as defined in (\ref{eq_h_implicit}). Condition 3 converts (\ref{eq_h_implicit}) to an implicit condition upon $\lambda_{s}(i, \beta_{m}, \alpha_{m})$ alone. Begin by differentiating (\ref{eq_h_implicit}) once for each $i$, $\beta_{m}$ and $\alpha_{m}$:

\begin{equation} \label{eq_triple_requirement}
    \frac{\partial^{3}\lambda_{s}}{\partial i \, \partial\beta_{m} \, \partial\alpha_{m}}=0
\end{equation}

Integrate (\ref{eq_triple_requirement}) over a cube with $\gamma_{1} \leq \beta_{m} \leq \gamma_{2} \leq i \leq \gamma_{3} \leq \alpha_{m} \leq \gamma_{4}$:

\begin{equation} \label{eq_cube_integration}
    \int_{\gamma_{1}}^{\gamma_{2}} \int_{\gamma_{2}}^{\gamma_{3}} \int_{\gamma_{3}}^{\gamma_{4}} \frac{\partial^{3}\lambda_{s}}{\partial i \, \partial\beta_{m} \, \partial\alpha_{m}} \, d\alpha_{m}  \, di \, d\beta_{m}   = 0
\end{equation}

The integral term of (\ref{eq_cube_integration}) is evaluated and simplified with the help of Condition 1:

\begin{equation} \label{eq_condition_3}
    \lambda_{s}(\gamma_{2}, \gamma_{1}, \gamma_{3}) - \lambda_{s}(\gamma_{2}, \gamma_{1}, \gamma_{4}) + \lambda_{s}(\gamma_{3}, \gamma_{1}, \gamma_{4}) - \lambda_{s}(\gamma_{3}, \gamma_{2}, \gamma_{4}) = 0
\end{equation}

Condition 3 requires (\ref{eq_condition_3}) to hold for all $\gamma_{1} \leq \gamma_{2} \leq \gamma_{3} \leq \gamma_{4}$. This condition can also be shown to be necessary and sufficient when the derivative of (\ref{eq_triple_requirement}) does not exist, as third order differentiability is not guaranteed by Condition 2.

\subsection{Explicit Construction of Shape Function}
\label{sec_explicit_h}

An explicit formula for $h(\beta_{m}, \alpha_{m})$ is obtained from (\ref{eq_h_implicit}) by recognising the close similarity between $\lambda_{s}(i, \beta_{m}, \alpha_{m})$ and $h(\beta_{m}, \alpha_{m})$. To see this similarity, differentiate (\ref{eq_h_implicit}) with only two of $i$, $\beta_{m}$ and $\alpha_{m}$ at a time:

\begin{itemize}
    \item[1.] Differentiating with $i$ and $\beta_{m}$
    
        $$ \frac{\partial^{2}\lambda_{s}}{\partial i \partial \beta_{m}} = \left. \frac{\partial^{2}h}{\partial \beta_{m} \partial \alpha_{m}} \right|_{\alpha_{m} = i}$$

        implies $\lambda_{s}(i, \beta_{m}, \alpha_{m}) \sim h(\beta_{m}, i)$ for constant $\alpha_{m}$.

    \item[2.] Differentiating with $i$ and $\alpha_{m}$

        $$ \frac{\partial^{2}\lambda_{s}}{\partial i \partial \alpha_{m}} = \left. \frac{\partial^{2}h}{\partial \beta_{m} \partial \alpha_{m}} \right|_{\beta_{m} = i}$$

        implies $\lambda_{s}(i, \beta_{m}, \alpha_{m}) \sim h(i, \alpha_{m})$ for constant $\beta_{m}$.

    \item[3.] Differentiating with $\beta_{m}$ and $\alpha_{m}$    

        $$ \frac{\partial^{2}\lambda_{s}}{\partial \beta_{m} \partial \alpha_{m}} = - \frac{\partial^{2}h}{\partial \beta_{m} \partial \alpha_{m}}$$

        implies $\lambda_{s}(i, \beta_{m}, \alpha_{m}) \sim -h(\beta_{m}, \alpha_{m})$ for constant $i$.
        
\end{itemize}

Case 1 could imply $h(\beta_{m}, \alpha_{m}) = \lambda_{s}(\alpha_{m}, \beta_{m}, \gamma)$ for a constant value of $\gamma$. This would only partly define $h(\beta_{m}, \alpha_{m})$ in the region of $\beta_{m} \leq \alpha_{m} \leq \gamma$. The constant $\gamma$ could be increased out to infinity, or some value that does not need to be practically exceeded. However, the general approach is to construct $h(\beta_{m}, \alpha_{m})$ by stitching all three cases together:

\begin{equation} \label{eq_h_explicit}
h(\beta_{m}, \alpha_{m}) = q(\alpha_{m}) - q(\beta_{m}) + \left\{\begin{array}{ll}
\lambda_{s}(\alpha_{m}, \beta_{m}, \gamma) & \beta_{m} \leq \alpha_{m} \leq \gamma \\
-\lambda_{s}(\gamma, \beta_{m}, \alpha_{m}) & \beta_{m} \leq \gamma \leq \alpha_{m} \\
\lambda_{s}(\beta_{m}, \gamma, \alpha_{m}) & \gamma \leq \beta_{m} \leq \alpha_{m}
\end{array}\right.
\end{equation}

\noindent where $\gamma$ can be any constant value, i.e. a current. The shape function is not unique because any function $q(i)$ that is differentiable can be chosen. Furthermore, $q(i)$ has no impact on the output of the time-periodic Preisach model $\lambda(t)$, and can be chosen for convenience. For example, requiring $\eta(\alpha)=0$ in (\ref{eq_general_preisach_weight_formulas}).

The explicit formula is verified by showing (\ref{eq_h_implicit}) is true solely by (\ref{eq_h_explicit}) and Condition 3 in (\ref{eq_condition_3}). Also, the differentiability of $h(\beta_{m}, \alpha_{m})$ along the boundaries $\gamma=\beta_{m}$ and $\gamma=\alpha_{m}$ can be verified with the help of Condition 1 and 2. This completes the construction of the time-periodic Preisach model with perfect test data. Section \ref{sec_fitting_reduced} constructs the model with symmetric test data.

\section{Method for Fitting Symmetric Test Results to Preisach Model}
\label{sec_fitting_reduced}

The previous section presented the time-periodic Preisach model assuming that the test data, $\lambda_{r}(i, \beta_{m}, \alpha_{m})$ and $\lambda_{d}(i, \beta_{m}, \alpha_{m})$, meet the requirements of Conditions 1 to 3 in Section \ref{sec_time_periodic_preisach}. Conditions 1 and 2 are easily satisfied, but there is no guarantee that the test coil will satisfy the congruency property of Condition 3. There are three broad approaches to forcing the congruency property. 1) Optimally fit test data to a reduced order model of $h(\beta_{m}, \alpha_{m})$ that has a finite number of parameters \cite{Zare2008, Hussain2018}. 2) Modify $\lambda_{s}(i, \beta_{m}, \alpha_{m})$ to satisfy Condition 3 and minimise error between the model output and test results. A considerable amount of testing is required as the minimum current $\beta_{m}$ and maximum current $\alpha_{m}$ are individually varied. 3) Reduce the number of tests to the ones that have to be exactly replicated while $h(\beta_{m}, \alpha_{m})$ remains under or exactly defined. Any missing information to the shape function can be filled in based on physical principles. This section explores the third option.

A simpler testing regime reduces the number of tests so that only one variable $\gamma_{m}$ is varied, where $\beta_{m}=-\gamma_{m}$ and $\alpha_{m}=\gamma_{m}$. This type of testing is called symmetric. The results of this approach are insufficient to fully describe $h(\beta_{m}, \alpha_{m})$, so the remaining information is inserted to best satisfy the internal minor loops condition. The symmetric test results are:

\begin{equation} \label{eq_reduced_r_def}
    \lambda_{r}(i, \gamma_{m}) = \lambda_{r}(i, -\gamma_{m}, \gamma_{m})
\end{equation}
\begin{equation} \label{eq_reduced_d_def}
    \lambda_{d}(i, \gamma_{m}) = \lambda_{d}(i, -\gamma_{m}, \gamma_{m})
\end{equation}

\noindent where $-\gamma_{m} \leq i \leq \gamma_{m}$ and $\gamma_{m} \geq 0$. The symmetric test results are split into differential and common mode components:

\begin{equation} \label{eq_reduced_diff_conversion}
    \lambda_{s}(i, \gamma_{m}) = \frac{1}{2}\big(\lambda_{d}(i, \gamma_{m}) - \lambda_{r}(i, \gamma_{m}) \big)
\end{equation}
\begin{equation} \label{eq_reduced_common_conversion}
    \lambda_{c}(i, \gamma_{m}) = \frac{1}{2}\big(\lambda_{d}(i, \gamma_{m}) + \lambda_{r}(i, \gamma_{m}) \big)
\end{equation}

The construction of $h(\beta_{m}, \alpha_{m})$ begins with substituting $\beta_{m}=-\gamma_{m}$ and $\alpha_{m}=\gamma_{m}$ into (\ref{eq_h_implicit}):

\begin{equation} \label{eq_h_implicit_reduced}
    \lambda_{s}(i, \gamma_{m}) = h(-\gamma_{m}, i) + h(i, \gamma_{m}) - h(-\gamma_{m}, \gamma_{m})
\end{equation}

Differentiating (\ref{eq_h_implicit_reduced}) with respect to both $i$ and $\gamma_{m}$:

\begin{equation}
    \frac{\partial^{2} \lambda_{s}}{\partial i \partial \gamma_{m}} = - \left. \frac{\partial^{2} h}{\partial \beta_{m} \partial \alpha_{m}} \right|_{\substack{\beta_{m} = -\gamma_{m} \\ \alpha_{m} = i}} + \left. \frac{\partial^{2} h}{\partial \beta_{m} \partial \alpha_{m}} \right|_{\substack{\beta_{m} = i \\ \alpha_{m} = \gamma_{m}}}
\end{equation}

This equation gives the impression that $\lambda_{s}(i, \gamma_{m})$ is built from two separate halves of $h(\beta_{m}, \alpha_{m})$. The first half is where $-\beta_{m} \leq \alpha_{m}$ and the other half is where $-\beta_{m} \geq \alpha_{m}$. Conversely, $h(\beta_{m}, \alpha_{m})$ can be built by splitting $\lambda_{s}(i, \gamma_{m})$ between these two halves of $h(\beta_{m}, \alpha_{m})$, but the exact split is unknown:

\begin{equation} \label{eq_h_explicit_reduced}   
    h(\beta_{m}, \alpha_{m}) = \frac{1}{2}\left\{ \begin{array}{ll} 
        \lambda_{s}(\alpha_{m}, -\beta_{m}) - d(\alpha_{m}, -\beta_{m}) & \quad -\beta_{m} \geq \alpha_{m} \\
        & \\
        \lambda_{s}(\beta_{m}, \alpha_{m}) + d(\beta_{m}, \alpha_{m}) & \quad -\beta_{m} \leq \alpha_{m}
    \end{array}\right.
\end{equation}

\noindent where $d(i, \gamma_{m})$ is the natural form of the splitting function $d$, which has the domain $-\gamma_{m} \leq i \leq \gamma_{m}$. The next step is to place boundary requirements on $d(i, \gamma_{m})$. The first is $h(i, i) = 0$:

\begin{equation} \label{eq_h_reduced_boundary1}
    h(i, i) = 0 = \frac{1}{2}\left\{ \begin{array}{ll} 
        \lambda_{s}(i, -i) - d(i, -i) & \quad i \leq 0 \\
        & \\
        \lambda_{s}(i, i) + d(i, i) & \quad i \geq 0
    \end{array}\right.
\end{equation}

Loop conditions from (\ref{eq_loop_bottom}) and (\ref{eq_loop_top}) require $\lambda_{s}(i, -i) = 0$ and $\lambda_{s}(i, i)=0$, which with (\ref{eq_h_reduced_boundary1}) implies $d(i, -i) = 0$ and $d(i, i) = 0$. Furthermore, this guarantees the continuity of $h(\beta_{m}, \alpha_{m})$ on the boundary between the two halves, $-\beta_{m} = \alpha_{m}$, and that $h(-i, i) = 0$. With this last result, (\ref{eq_h_explicit_reduced}) is in agreement with (\ref{eq_h_implicit_reduced}).

The second boundary requirement, because of Condition 2, is for the partial derivative of $h(\beta_{m}, \alpha_{m})$ with respect to $\beta_{m}$ to be continuous on the line $-\beta_{m} = \alpha_{m}$. Differentiating $h(\beta_{m}, \alpha_{m})$ with respect to $\beta_{m}$ gives:

\begin{equation} \label{eq_h_explicit_betam}   
    \frac{\partial h}{\partial \beta_{m}} = \frac{1}{2}\left\{ \begin{array}{ll} 
        -\left. \frac{\partial \lambda_{s}}{\partial \gamma_{m}} \right|_{\substack{i = \alpha_{m} \\ \gamma_{m} = -\beta_{m}}} + \left. \frac{\partial d}{\partial \gamma_{m}} \right|_{\substack{i = \alpha_{m} \\ \gamma_{m} = -\beta_{m}}} & \quad -\beta_{m} \geq \alpha_{m} \\
        & \\
        \quad \left. \frac{\partial \lambda_{s}}{\partial i} \right|_{\substack{i = \beta_{m} \\ \gamma_{m} = \alpha_{m}}} + \left. \frac{\partial d}{\partial i}\right|_{\substack{i = \beta_{m} \\ \gamma_{m} = \alpha_{m}}} & \quad -\beta_{m} \leq \alpha_{m}
    \end{array}\right.
\end{equation}

The notation of (\ref{eq_h_explicit_betam}) is cumbersome, so the following notational simplification is made to express the substitution of arguments after differentiation:

$$ \left. \frac{\partial \lambda_{s}}{\partial \gamma_{m}} \right|_{\substack{i = \alpha_{m} \\ \gamma_{m} = -\beta_{m}}}  \rightarrow \frac{\partial \lambda_{s}}{\partial \gamma_{m}}(\alpha_{m}, -\beta_{m}) $$

Requiring the derivatives of each half of (\ref{eq_h_explicit_betam}) to be equal at the boundary $\beta_{m} = -\gamma_{m}$ and $\alpha_{m} = \gamma_{m}$ results in:

\begin{equation} \label{eq_h_explicit_betam_step2}
    -\frac{1}{2}\frac{\partial \lambda_{s}}{\partial \gamma_{m}}(\gamma_{m}, \gamma_{m}) + \frac{1}{2}\frac{\partial d}{\partial \gamma_{m}}(\gamma_{m}, \gamma_{m}) = \frac{1}{2}\frac{\partial \lambda_{s}}{\partial i}(-\gamma_{m}, \gamma_{m}) + \frac{1}{2}\frac{\partial d}{\partial i}(-\gamma_{m}, \gamma_{m})
\end{equation}

\noindent which can be adjusted to keep all partial derivatives with respect to $i$:

\begin{equation}\label{eq_h_explicit_betam_step3}
    \frac{1}{2}\frac{\partial \lambda_{s}}{\partial i}(\gamma_{m}, \gamma_{m}) - \frac{1}{2}\frac{\partial \lambda_{s}}{\partial i}(-\gamma_{m}, \gamma_{m}) = \frac{1}{2}\frac{\partial d}{\partial i}(\gamma_{m}, \gamma_{m}) + \frac{1}{2}\frac{\partial d}{\partial i}(-\gamma_{m}, \gamma_{m})
\end{equation}

The adjustment is possible because $\lambda_{s}(\gamma_{m}, \gamma_{m}) = 0$ and $d(\gamma_{m}, \gamma_{m})=0$, which are differentiated with respect to $\gamma_{m}$ to obtain the replacement formulas for $\partial \lambda_{s}/\partial \gamma_{m}$ and $\partial d/\partial \gamma_{m}$ in (\ref{eq_h_explicit_betam_step2}). Further simplification of (\ref{eq_h_explicit_betam_step3}) requires splitting $\lambda_{s}$ and $d$ into odd and even components:

$$ \lambda_{se}(i, \gamma_{m}) = \frac{1}{2} \Big(\lambda_{s}(i, \gamma_{m}) + \lambda_{s}(-i, \gamma_{m}) \Big) $$ 
$$ \lambda_{so}(i, \gamma_{m}) = \frac{1}{2} \Big(\lambda_{s}(i, \gamma_{m}) - \lambda_{s}(-i, \gamma_{m}) \Big) $$
$$ d_{e}(i, \gamma_{m}) = \frac{1}{2} \Big(d(i, \gamma_{m}) + d(-i, \gamma_{m}) \Big) $$

\begin{equation} \label{eq_do_def}
    d_{o}(i, \gamma_{m}) = \frac{1}{2} \Big(d(i, \gamma_{m}) - d(-i, \gamma_{m}) \Big)
\end{equation} 

The final form of the $-\beta_{m} = \alpha_{m}$ boundary differentiability condition becomes:

\begin{equation}\label{eq_h_explicit_betam_step4}
    \frac{\partial \lambda_{se}}{\partial i}(\gamma_{m}, \gamma_{m}) = \frac{\partial d_{o}}{\partial i}(\gamma_{m}, \gamma_{m})
\end{equation}

This completes the requirement for Condition 2 with respect to the derivative of $\beta_{m}$. The condition for $\partial h / \partial \alpha_{m}$ to be continuous on the boundary $-\beta_{m} = \alpha_{m}$ is identical to (\ref{eq_h_explicit_betam_step4}). This is because $h(-\gamma_{m}, \gamma_{m}) = 0$, which after differentiating with respect to $\gamma_{m}$ gives:

\begin{equation}
    -\frac{\partial h}{\partial \beta_{m}}(-\gamma_{m}, \gamma_{m}) + \frac{\partial h}{\partial \alpha_{m}}(-\gamma_{m}, \gamma_{m}) = 0
\end{equation}

\noindent and states both derivatives are equal along the boundary. This completes all necessary requirements for $d(i, \gamma_{m})$. An optional requirement is for all second order partial derivatives to be continuous on the line $-\beta_{m} = \alpha_{m}$:

\begin{equation} \label{eq_d2_continuity_condition}
    \frac{\partial^{2} \lambda_{so}}{\partial i \partial \gamma_{m}}(\gamma_{m}, \gamma_{m}) = \frac{\partial^{2} d_{e}}{\partial i \partial \gamma_{m}}(\gamma_{m}, \gamma_{m})
\end{equation}

Further specification of $d(i, \gamma_{m})$ is based on desired hysteresis properties.

\subsection{Internal Minor Loops Condition}
\label{sec_internal_minor}

The internal minor loops condition states that all minor loops should remain within their major loop, as demonstrated in Fig. \ref{fig_major_minor_diff}. Also, any child minor loops should remain within their parent minor loop. This condition is consistent with the idea that magnetic remanence and coercivity only increase with greater applied magnetic field. This section derives the internal minor loop condition for time-periodic Preisach models. Furthermore, an equation determining the feasibility of satisfying both the internal minor loop condition and (\ref{eq_h_explicit_betam_step4}) is stated.

\begin{figure}
	\centering
		\includegraphics[scale=1]{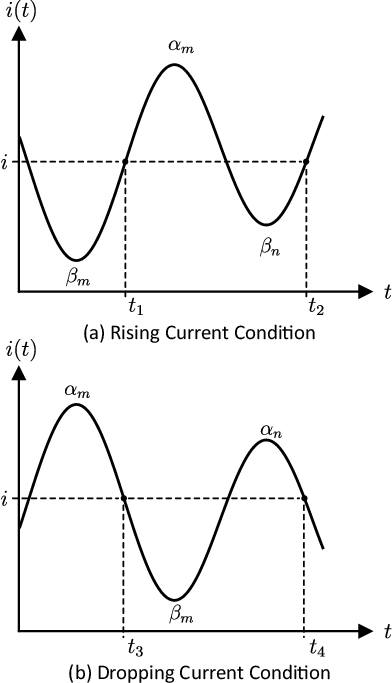}
	\caption{Test current waveforms to specify internal minor loops condition.}
	\label{fig_internal_meaning}
\end{figure}

The first step is to specify the orientation of the major loop. For magnetic hysteresis, the positive orientation (anti-clockwise) is required with $\lambda_{s}(i, \beta_{m}, \alpha_{m}) \geq 0$ and  $\lambda_{d}(i, \beta_{m}, \alpha_{m}) \geq \lambda_{r}(i, \beta_{m}, \alpha_{m})$. The internal minor loops condition is satisfied if from Fig. \ref{fig_internal_meaning}(a) that $\lambda(t_{2}) \geq \lambda(t_{1})$ for all $\beta_{m} \leq \beta_{n} \leq i \leq \alpha_{m}$, and from Fig. \ref{fig_internal_meaning}(b) that $\lambda(t_{4}) \leq \lambda(t_{3})$ for all $\beta_{m} \leq i \leq \alpha_{n} \leq \alpha_{m}$.

The internal minor loop condition can be reduced to $\mu(\beta, \alpha) \geq 0$. This derivation begins with evaluating $\lambda(t_{2}) - \lambda(t_{1})$ according to (\ref{eq_preisach_characteristic}):

\begin{equation} \label{eq_internal_step1}
    \lambda(t_{2}) - \lambda(t_{1}) = 2\Big(h(\beta_{n}, \alpha_{m}) - h(\beta_{m}, \alpha_{m}) - h(\beta_{n}, i) + h(\beta_{m}, i) \Big)
\end{equation}

\begin{figure}
	\centering
		\includegraphics[scale=1]{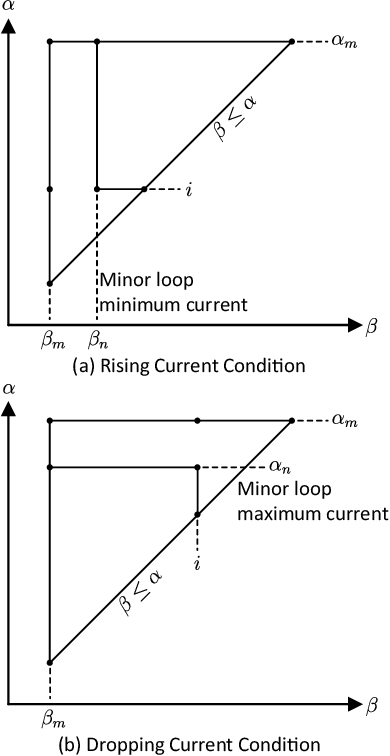}
	\caption{Spilt of the domain of $p(\beta, \alpha)$ into $S^{+}(t_{2})$ and $S^{-}(t_{2})$ for (a), and $S^{+}(t_{4})$ and $S^{-}(t_{4})$ for (b), according to the times shown in Fig. \ref{fig_internal_meaning}.}
	\label{fig_internal_construction}
\end{figure}

The right hand side of (\ref{eq_internal_step1}) is identifiable as the integral of a rectangular region in $p(\beta, \alpha)$:

\begin{equation} \label{eq_internal_step2}
    \lambda(t_{2}) - \lambda(t_{1}) = 2\int_{i}^{\alpha_{m}} \int_{\beta_{m}}^{\beta_{n}} \frac{\partial^{2}h}{\partial \beta_{m} \partial \alpha_{m}}(\beta, \alpha)\, d\!\beta \, d\!\alpha  = 2\int_{i}^{\alpha_{m}} \int_{\beta_{m}}^{\beta_{n}} \mu(\beta, \alpha)\, d\!\beta \, d\!\alpha 
\end{equation}

\noindent which can be checked by analysing the difference in $S^{+}(t)$ and $S^{-}(t)$ between time $t_{1}$ and $t_{2}$ in Fig. \ref{fig_internal_construction}(a). Therefore, since the internal minor loops condition requires $\lambda(t_{2}) - \lambda(t_{1}) \geq 0$ for all $\beta_{m} \leq \beta_{n} \leq i \leq \alpha_{m}$, (\ref{eq_internal_step2}) implies $\mu(\beta, \alpha) \geq 0$ and $\lambda(t_{4}) \leq \lambda(t_{3})$ through similar working. Following sections will create methods for making $\mu(\beta, \alpha) \geq 0$.

The feasibility of satisfying the internal minor loops condition is dependent on the test data $\lambda_{s}$. Appendix \ref{sec_generating_method} derives a function $m_{0}(\gamma_{m})$ from $\lambda_{s}(i, \gamma_{m})$ so that if $m_{0}(\gamma_{m}) \geq 0$ then it is possible to choose $d(i, \gamma)$ with $\mu(\beta, \alpha) \geq 0$ and for (\ref{eq_h_explicit_betam_step4}) to be true. The function $m_{0}(\gamma_{m})$ is as follows:

\begin{equation} \label{eq_g_requirement_1_short}
     m_{0}(\gamma_{m}) = -\frac{1}{2} \int_{-\gamma_{m}}^{\gamma_{m}} \left| \frac{\partial^{2}\lambda_{s}}{\partial i \partial \gamma_{m}}(\beta, \gamma_{m}) \right| d\!\beta - \frac{\partial \lambda_{se}}{\partial i}(\gamma_{m}, \gamma_{m})
\end{equation}

\subsection{Approximate Method of Constructing Splitting Function}
\label{sec_approximate_method}

\begin{figure}
	\centering
		\includegraphics[scale=1]{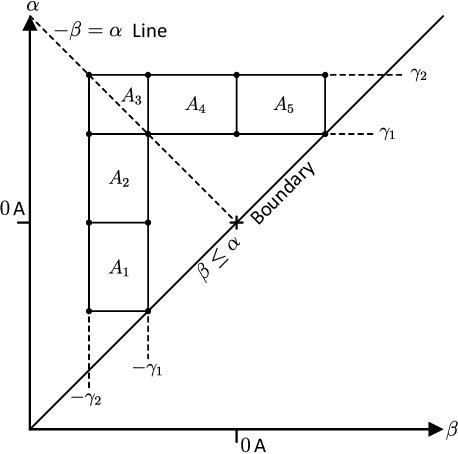}
	\caption{Regions in the domain of $\mu(\beta, \alpha)$ for deriving necessary conditions for internal minor loops.}
	\label{fig_approximate_fit}
\end{figure}

Appendix \ref{sec_generating_method} in developing (\ref{eq_g_requirement_1_short}) also presents a general method of constructing the splitting function $d(i, \gamma_{m})$ to satisfy $\mu(\beta, \alpha) \geq 0$. In practice, this method is difficult to implement so an approximation is developed here instead. The approximate method creates $d(i, \gamma_{m})$ by requiring the integral of $\mu(\beta, \alpha)$ over the areas $A_{1}$ to $A_{5}$ in Fig. \ref{fig_approximate_fit} to be positive for all $0 \leq \gamma_{1} \leq \gamma_{2}$. These requirements form a set of necessary conditions for $\mu(\beta, \alpha) \geq 0$, but are insufficient to guarantee $\mu(\beta, \alpha) \geq 0$ everywhere, as not all possible areas of the $(\beta, \alpha)$ plane are specified.

The construction of $d(i, \gamma_{m})$ begins by integrating over $A_{j}$ for $j\in\{1,2,3,4,5\}$:

\begin{equation} \label{eq_M_def}
    M_{j} = \iint_{A_{j}} \mu(\beta, \alpha) \, d\beta \, d\alpha
\end{equation}

The requirement for all $M_{j} \geq 0$ and (\ref{eq_h_explicit_betam_step4}) encourages the following construction of $d(i, \gamma_{m})$:

\begin{equation} \label{eq_d_approx}
    d(i, \gamma_{m}) = m\left(\frac{i}{\gamma_{m}} \right)\lambda_{s}(i, \gamma_{m})
\end{equation}

\noindent where $m(\zeta)$ is an odd function with $\zeta = i / \gamma_{m}$, $m(\zeta) = m(-\zeta)$, and is separated into a left and right half:

\begin{equation} \label{eq_m_half}
    m(\zeta) = \left\{ \begin{array}{ll}
         \hat{m}(-\zeta) & \zeta < 0 \\
         \hat{m}(\zeta) & \zeta \geq 0
    \end{array} \right.
\end{equation}

\noindent with $\hat{m}(\zeta)$ being defined on the domain $0 \leq \zeta \leq 1$. The justification for this choice is given in Appendix \ref{sec_approximate_measures} and is summarised here with conditions upon $\lambda_{s}(i, \gamma_{m})$ and $\hat{m}(\zeta)$ in the list below. Each condition assumes the conditions above it:

\begin{itemize}
    \item[1)] For $(M_{1} + M_{2})/2 + M_{3} \geq 0$ and $M_{3} + (M_{4} + M_{5})/2 \geq 0$ requires $\lambda_{se}(i, \gamma_{m}) \geq  |\lambda_{so}(i, \gamma_{m})| \geq 0$
    \item[2)] $\hat{m}(0)=0$ for $m(\zeta)$ to be an odd function.
    \item[3)] $\hat{m}(1)=1$ for (\ref{eq_h_explicit_betam_step4}) to be true.
    \item[4)] $d\hat{m}(1)/d\zeta = 0$ for (\ref{eq_d2_continuity_condition}) to be true.
    \item[5)] For $M_{3} \geq 0$ requires $\hat{m}(\zeta) \leq 1$.
    \item[6)] For $M_{1} + M_{2} \geq 0$ and $M_{4} + M_{5} \geq 0$ requires:

            \begin{equation} \label{eq_m1_measure}
                \hat{m}(\zeta) \geq m_{1}(\zeta) = \max_{\gamma \geq 0} \left\{ \frac{|\lambda_{so}(\zeta \gamma, \gamma)|}{\lambda_{se}(\zeta \gamma, \gamma)} \right\} \geq 0
            \end{equation}

            Also $m_{1}(\zeta) \leq 1$ is equivalent to the first condition.
            
    \item[6)] For $M_{1} \geq 0$ and $M_{4} \geq 0$ requires:

            \begin{equation} \label{eq_m2_measure}
                \hat{m}(\zeta) \geq m_{2}(\zeta) = \max_{\gamma \geq 0} \left\{ \frac{|\lambda_{SE}(\zeta \gamma, \gamma) -\lambda_{so}(\zeta \gamma, \gamma)|}{\lambda_{se}(\zeta \gamma, \gamma) - \lambda_{so}(\zeta \gamma, \gamma)} \right\}
            \end{equation}
            
            where $\lambda_{SE}(\gamma_{1}, \gamma_{2}) = \lambda_{se}(\gamma_{1}, \gamma_{2}) + \lambda_{se}(0, \gamma_{1}) -  \lambda_{se}(0, \gamma_{2})$.

    \item[7)] For $M_{2} \geq 0$ and $M_{5} \geq 0$ requires:

            \begin{equation} \label{eq_m3_measure}
                \hat{m}(\zeta) \geq m_{3}(\zeta) = \max_{\gamma \geq 0} \left\{ \frac{|\lambda_{SE}(\zeta \gamma, \gamma) + \lambda_{so}(\zeta \gamma, \gamma)|}{\lambda_{se}(\zeta \gamma, \gamma) + \lambda_{so}(\zeta \gamma, \gamma)} \right\}
            \end{equation}

\end{itemize}

These conditions places a series of bounds upon $\hat{m}(\zeta)$, which are achievable if $m_{1}(\zeta) \leq 1$, $m_{2}(\zeta) \leq 1$ and $m_{3}(\zeta) \leq 1$. A function that satisfies these conditions is:

\begin{equation} \label{eq_mhat_choice}
    \hat{m}(\zeta) = \frac{1 - a\zeta e^{-a} - e^{-a\zeta}}{1 - ae^{-a} - e^{-a}}
\end{equation}

\noindent where $a$ is a parameter to be selected. The closer $a \rightarrow \infty$, the more $\hat{m}(\zeta) \rightarrow 1$ on $\zeta \in (0,1]$. The last step of this section is to update $h(\beta_{m}, \alpha_{m})$ in (\ref{eq_h_explicit_reduced}) with the approximate $d(i, \gamma_{m})$: 

\begin{equation} \label{eq_h_explicit_approximate}   
    h(\beta_{m}, \alpha_{m}) = \frac{1}{2}\left\{ \begin{array}{ll} 
        \left(1 - m\left(\frac{\alpha_{m}}{-\beta_{m}} \right)\right)\lambda_{s}(\alpha_{m}, -\beta_{m}) & \quad -\beta_{m} \geq \alpha_{m} \\
        & \\
        \left(1 + m\left(\frac{\beta_{m}}{\alpha_{m}} \right)\right)\lambda_{s}(\beta_{m}, \alpha_{m}) & \quad -\beta_{m} \leq \alpha_{m}
    \end{array}\right.
\end{equation}

\subsection{Fitting the Common Mode Function}
\label{sec_common_mode}

Previous sections fitted symmetric test data to create the shape function $h(\beta_{m}, \alpha_{m})$. This section completes the time-periodic Preisach model with an appropriate fit for $c(i(t), \beta_{m}, \alpha_{m})$. Two criteria govern selection: 1) $\lambda(t)$ should trace loops within the vicinity of the centre line, 2) Test data should be exactly replicable. The centre line is defined as the greatest extent of the odd component of $\lambda_{c}$:

\begin{equation} \label{eq_center_line_formula}
    \lambda_{cl}(i) = \frac{1}{2}\left(\lambda_{c}(i, \gamma_{M}) - \lambda_{c}(-i, \gamma_{M})\right)
\end{equation}

\noindent where $\gamma_{M} = \max\{\gamma_{m}\}$ is the maximum tested peak current. Criteria 2 requires the perturbation between $\lambda_{c}$ and $\lambda_{cl}$:

\begin{equation}
    \lambda_{\Delta c}(i, \gamma_{m}) = \lambda_{c}(i, \gamma_{m}) - \lambda_{cl}(i)
\end{equation}

Therefore, the fitted common mode function is:

\begin{equation} \label{eq_common_mode_function_impl}
    c(i, \beta_{m}, \alpha_{m}) = \lambda_{cl}(i) + \lambda_{\Delta c}(i - (\beta_{m} + \alpha_{m})/2, (\alpha_{m} - \beta_{m})/2) + h(\beta_{m}, i) - h(i, \alpha_{m})   
\end{equation}

\noindent so that $\lambda_{c}(i, \gamma_{m}) = \lambda_{c}(i, -\gamma_{m}, \gamma_{m})$.

\section{Harmonic Domain Model for Transformer Open-Circuit Characteristics}
\label{sec_preisach_harmonic_domain}

This section develops a harmonic domain model, through the creation of a frequency coupling matrix, for a transformer's open-circuit test characteristics. The time-periodic Preisach model does not give an explicit operator of the form, $i(t)=f\{v(t)\}$, which is required to directly construct the harmonic domain model. Rather, the governing equations are $\lambda(t) = H_{t}\{i(t)\}$ and $v(t)=d\lambda(t)/dt$, which are inverted in the harmonic domain to give the frequency coupling matrix. This process is separated into the following steps:

\begin{itemize}
    \item[1.] Construct the frequency coupling matrices for $\lambda(t) = H_{t}\{i(t)\}$ to give $\Delta\boldsymbol{\Lambda} = P^{(1)}\Delta\mathbf{I} + P^{(2)}\overline{\Delta\mathbf{I}}$, where $\Delta\boldsymbol{\Lambda} = \boldsymbol{\Lambda} - \boldsymbol{\Lambda}_{B}$ with:

    \begin{equation} \label{eq_lambda_coefficients}
        \boldsymbol{\Lambda} = \begin{bmatrix} \Lambda_{0}, \Lambda_{1}, \Lambda_{2}, \cdots \end{bmatrix}^{T} \; \; \text{and} \; \; \Lambda_{n}=\frac{2-\delta[n]}{T} \int_{0}^{T}\lambda(t)e^{-jn\omega t}dt
    \end{equation}

    Similarly, $\boldsymbol{\Lambda}_{B}$ are the harmonic coefficients of the base equivalent flux linkage $\lambda_{B}(t)$, where $\lambda_{B}(t) = H_{t}\{i_{B}(t)\}$ and $v_{B}(t)=d\lambda_{B}(t)/dt$. The construction of $P^{(1)}$ and $P^{(2)}$ through the linearisation of $\lambda(t) = H_{t}\{i(t)\}$ is explained in Section \ref{sec_linear_preisach}.

    \item[2. ] Invert the system of linear equations $\Delta\boldsymbol{\Lambda} = P^{(1)}\Delta\mathbf{I} + P^{(2)}\overline{\Delta\mathbf{I}}$ to give $\Delta\mathbf{I} = F^{(1)}\Delta\boldsymbol{\Lambda} + F^{(2)}\overline{\Delta\boldsymbol{\Lambda}}$. Both the $P^{(1)}$ and $P^{(2)}$ matrices have infinite extent; in practice, these matrices are truncated before performing the inversion. The inversion also requires splitting both $\Delta\boldsymbol{\Lambda}$ and $\Delta\mathbf{I}$ into real and imaginary components.

    \item[3. ] The voltage equation $v(t)=d\lambda(t)/dt$ is converted into the harmonic domain $\mathbf{V} = D\boldsymbol{\Delta\boldsymbol{\Lambda}}$, where $D=\text{diag}(\begin{bmatrix} 0, j\omega, j2\omega, j3\omega, \cdots \end{bmatrix})$. Differential operator $D$ is inverted to give the integral operator $D^{-1}$. Special consideration is given to the dc component of $D$, i.e. the first component, for it to be invertible. In practice, the first component of $D^{-1}$ is treated as $\infty$.

    \item[4. ] Substitute $\boldsymbol{\Delta\boldsymbol{\Lambda}} = D^{-1}\mathbf{V}$ into $\Delta\mathbf{I} = F^{(1)}\Delta\boldsymbol{\Lambda} + F^{(2)}\overline{\Delta\boldsymbol{\Lambda}}$ to give $\Delta\mathbf{I} = Y^{(1)}\Delta\mathbf{V} + Y^{(2)}\overline{\Delta\mathbf{V}}$. The admittance matrices are formed as $Y^{(1)}=F^{(1)}D^{-1}$ and $Y^{(2)}=F^{(2)}\overline{D^{-1}}$.
        
\end{itemize}

\subsection{Linearisation of the Time-Periodic Preisach Operator}
\label{sec_linear_preisach}

Linearisation in the harmonic domain begins by substituting $\lambda(t) = H_{t}\{i(t)\}$ into (\ref{eq_lambda_coefficients}), and then substituting in the harmonic phasor series for $i(t)$:

\begin{equation} \label{eq_lambda_coeff}
    \Lambda_{n}(\mathbf{I})=\frac{2-\delta[n]}{T} \int_{0}^{T}H_{t}\Bigg\{\Re \Bigg\{\sum_{m=0}^{\infty}I_{m}e^{jm \omega t} \Bigg\}\Bigg\}e^{-jn\omega t}dt
\end{equation}

The linearisation requires differentiating $\Lambda_{n}$ with $I_{m}$. It is expected that the complex derivative $\partial \Lambda_{n}/\partial I_{m}$ does not exist. Therefore, $\Lambda_{n}$ and $I_{m}$ are separated into real and imaginary components. The separation of (\ref{eq_lambda_coeff}) into $\Lambda_{n}^{\Re}$ and $\Lambda_{n}^{\Im}$ gives the following two equations:

\begin{equation} \label{eq_lambda_coeff_real}
    \Lambda_{n}^{\Re}(\mathbf{I})=\frac{2-\delta[n]}{T} \int_{0}^{T}H_{t}\Bigg\{\Re \Bigg\{\sum_{m=0}^{\infty}I_{m}e^{jm \omega t} \Bigg\}\Bigg\}\cos(jn\omega t)\,dt
\end{equation}

\begin{equation} \label{eq_lambda_coeff_imag}
    \Lambda_{n}^{\Im}(\mathbf{I})= -\frac{2-\delta[n]}{T} \int_{0}^{T}H_{t}\Bigg\{\Re \Bigg\{\sum_{m=0}^{\infty}I_{m}e^{jm \omega t} \Bigg\}\Bigg\}\sin(jn\omega t)\,dt
\end{equation}

Both $\Lambda_{n}^{\Re}$ and $\Lambda_{n}^{\Im}$ are differentiated with respect to $\Delta I_{m}^{\Re}$ and $\Delta I_{m}^{\Im}$. The symbol $\psi$ is used to represent both terms, not at the same time, so that $\psi = \Delta I_{m}^{\Re}$ or $\psi = \Delta I_{m}^{\Im}$ to simplify working:

\begin{equation} \label{eq_dlambda_coeff_real}
    \frac{\partial\Lambda_{n}^{\Re}}{\partial \psi} =\frac{2-\delta[n]}{T} \int_{0}^{T} \frac{\partial}{\partial \psi}H_{t}\Bigg\{\Re \Bigg\{\sum_{m=0}^{\infty}I_{m}e^{jm \omega t} \Bigg\}\Bigg\}\cos(jn\omega t)\,dt
\end{equation}

\begin{equation} \label{eq_dlambda_coeff_imag}
    \frac{\partial\Lambda_{n}^{\Im}}{\partial \psi} = -\frac{2-\delta[n]}{T} \int_{0}^{T}\frac{\partial}{\partial \psi}H_{t}\Bigg\{\Re \Bigg\{\sum_{m=0}^{\infty}I_{m}e^{jm \omega t} \Bigg\}\Bigg\}\sin(jn\omega t)\,dt
\end{equation}

The next step is to evaluate $\partial H_{t}/\partial \psi$ according to (\ref{eq_preisach_characteristic}). To avoid repeated working for $\psi = \Delta I_{m}^{\Re}$ and $\psi = \Delta I_{m}^{\Im}$, the differential term within (\ref{eq_dlambda_coeff_real}) and (\ref{eq_dlambda_coeff_imag}) is generalised to:

\begin{equation}
    \frac{\partial}{\partial \psi}H_{t}\Bigg\{\Re \Bigg\{\sum_{m=0}^{\infty}I_{m}e^{jm \omega t}\Bigg\}\Bigg\} = \frac{\partial}{\partial \psi}\left(H_{t}\left\{i(t, \psi) \right\}\right)
\end{equation}

\noindent where $i(t, \psi) = i_{B}(t) + \psi w(t)$, and $w(t) = \cos(jm \omega t)$ or $w(t) = -\sin(jm \omega t)$ depending on the selection of $\psi$. The partial derivatives will be evaluated at $\psi = 0$, so that the linearisation will be about the function $i_{B}(t)$. The shape function form of the time-periodic Preisach model (\ref{eq_preisach_characteristic}) becomes:

$$ H_{t}\{i(t, \psi)\} = c(i(t, \psi), \, \beta_{m}(\psi), \, \alpha_{m}(\psi)) \, + \, \epsilon\{\rho_{0}(\psi)\}h(\beta_{m}(\psi), \, \alpha_{m}(\psi)) $$
\begin{equation} \label{eq_preisach_characteristic_psi}
     + 2 \!\!\! \sum_{j = 1}^{N(t, \psi)-1} \epsilon\{\rho_{j}(\psi)\}\, h(\beta\{\rho_{j}(\psi)\}, \, \alpha\{\rho_{j}(\psi)\})
\end{equation}

The existence of $\partial H_{t}/\partial \psi$ is dependent on an appropriate selection for $i_{B}(t)$. This requires the function $\lambda(t, \psi) = H_{t}\left\{i(t, \psi) \right\}$ to be smooth in the $\psi$ direction for all possible $w(t)$. The derivative $\partial H_{t}/\partial \psi$ does not need to exist at every time $t\in[0,T]$, but the integration of (\ref{eq_dlambda_coeff_real}) and (\ref{eq_dlambda_coeff_imag}) has to be possible. It is impractical to discuss all possible choices for $i_{B}(t)$ in this work. Therefore, the following basic conditions are assumed about $i_{B}(t)$ and $i(t, \psi)$ for $\psi$ in an open interval $\psi \in (-\kappa, \kappa)$ for some $\kappa > 0$:

\begin{itemize}
    \item[a) ] $i_{B}(t)$ is differentiable over $t \in [0, T]$.
    \item[b) ] All historical minimums $\beta_{k}(\psi)$ and maximums $\alpha_{l}(\psi)$ occur at distinct times, $t_{k}(\psi)$ and $t_{l}(\psi)$ respectively, that are a function of $\psi$:

    \begin{equation} \label{eq_beta_k_psi}
        \beta_{k}(\psi) = i_{B}(t_{k}(\psi)) + \psi w(t_{k}(\psi))
    \end{equation}
    \begin{equation} \label{eq_alpha_l_psi}
        \alpha_{l}(\psi) = i_{B}(t_{l}(\psi)) + \psi w(t_{l}(\psi))
    \end{equation}

    where $\beta_{k}(\psi)$ and $\alpha_{l}(\psi)$ are differentiable over $\psi \in (-\kappa, \kappa)$.
    
    \item[c) ] Historical minimums and maximums do not come in to and out of existence over the domain of $\psi \in (-\kappa, \kappa)$, but can be created or wiped over time $t\in[0,T]$.

    \item[d) ] The global minimum $\beta_{m}(\psi)$ and the global maximum $\alpha_{m}(\psi)$ each occur at only one distinct time in $t \in [0, T)$.
\end{itemize}

The above conditions can guarantee $\partial H_{t}/\partial \psi$ to exist for all time $t \in [0, T]$, except at points in time where the historical minimums and maximums of $i_{B}(t)$ are wiped out. These are times when $N(t, \psi)$ makes some of its discrete changes. The partial derivative $\partial H_{t}/\partial \psi$ is built up in steps by calculating the partial derivatives of its terms as follows:

\begin{itemize}
    \item[a) ] For $i(t, \psi)$:
    $$ \frac{\partial i(t, \psi)}{\partial \psi} = w(t)$$    

    \item[b) ] For $\beta_{k}(\psi)$ starting with (\ref{eq_beta_k_psi}):

    $$ \frac{d\beta_{k}}{d\psi} = \frac{di_{B}}{dt}\frac{dt_{k}}{d\psi} + w(t_{k}(\psi)) + \psi \frac{dw}{dt}\frac{dt_{k}}{d\psi}$$

    Evaluating this expression at $\psi = 0$:

    $$ \left. \frac{d\beta_{k}}{d\psi} \right|_{\psi = 0} = w(t_{k}(0))$$

    because for extrema $di_{B}/dt = 0$. Note, for the linearisation of $H_{t}$ about $i_{B}(t)$ it is not necessary to know what $t_{k}(\psi)$ is other than for $\psi = 0$.

    \item[c) ] Similarly for $\alpha_{l}(\psi)$ starting with (\ref{eq_alpha_l_psi}):

    $$ \left. \frac{d\alpha_{l}}{d\psi} \right|_{\psi = 0} = w(t_{l}(0))$$

    \item[d) ] By condition 3 above, $\epsilon\{\rho_{j}(\psi)\}$ is constant with respect to $\psi$, so $d\epsilon/d\psi = 0$.

    \item[e) ] For $h(\beta_{k}(\psi), \alpha_{l}(\psi))$, the chain rule is applied:

    $$ \frac{\partial h}{\partial \psi} = \frac{\partial h}{\partial \beta_{m}}\frac{d\beta_{k}}{d\psi} + \frac{\partial h}{\partial \alpha_{m}}\frac{d\alpha_{l}}{d\psi}$$

    Evaluating this at $\psi = 0$:
    
    $$ \left. \frac{\partial h}{\partial \psi} \right|_{\psi = 0} = w(t_{k}(0)) \frac{\partial h}{\partial \beta_{m}}(\beta_{k}(0), \alpha_{l}(0)) + w(t_{l}(0)) \frac{\partial h}{\partial \alpha_{m}}(\beta_{k}(0), \alpha_{l}(0))$$
    
    Simplify the above equation by defining the following operator:

    $$ \partial h\{\rho_{j}, \sigma_{j}\} = \beta\{\sigma_{j}\} \frac{\partial h}{\partial \beta_{m}} (\beta\{\rho_{j}\}, \alpha\{\rho_{j}\}) +  \alpha\{\sigma_{j}\}\frac{\partial h}{\partial \alpha_{m}} (\beta\{\rho_{j}\}, \alpha\{\rho_{j}\})$$

    where $\rho_{j}=(\beta_{k}(0), \alpha_{l}(0))$ and $\sigma_{j} = (w(t_{k}(0)), w(t_{l}(0)))$.

    \item[f) ] For $h(\beta_{k}(\psi), i(t, \psi))$ and $h(i(t, \psi), \alpha_{l}(\psi))$, which occur when $j = N(t, \psi) - 1$, a modification is required for $\rho_{j}$ and $\sigma_{j}$. 
    
    $$ h(\beta_{k}(\psi), i(t, \psi)) \text{  has  } \rho_{j} = (\beta_{k}(0), i_{B}(t)) \text{  and  } \sigma_{j} = (w(t_{k}(0)),w(t)) $$ 
    $$ h(i(t, \psi), \alpha_{l}(\psi)) \text{  has  } \rho_{j} = (i_{B}(t), \alpha_{l}(0)) \text{  and  } \sigma_{j} = (w(t), w(t_{l}(0))) $$

    \item[g) ] For $c(i(t, \psi), \, \beta_{m}(\psi), \, \alpha_{m}(\psi))$:

    $$\frac{\partial c}{\partial \psi} = \frac{\partial c}{\partial i} \frac{\partial i(t, \psi)}{\partial \psi} + \frac{\partial c}{\partial \beta_{m}}\frac{d\beta_{m}}{d\psi} + \frac{\partial c}{\partial \alpha_{m}}\frac{d\alpha_{m}}{d\psi} $$

    which is evaluated at $\psi = 0$ with the definition of the term $\partial c$:

    $$ \partial c\{i_{B}(t), w(t), \rho_{0}, \sigma_{0}\} = \left. \frac{\partial c}{\partial \psi} \right|_{\psi = 0} = w(t)\frac{\partial c}{\partial i}(i_{B}(t), \beta\{\rho_{0}\}, \alpha\{\rho_{0}\}) $$
    $$ + \beta\{\sigma_{0}\}\frac{\partial c}{\partial \beta_{m}}(i_{B}(t), \beta\{\rho_{0}\}, \alpha\{\rho_{0}\}) + \alpha\{\sigma_{0}\}\frac{\partial c}{\partial \alpha_{m}}(i_{B}(t), \beta\{\rho_{0}\}, \alpha\{\rho_{0}\}) $$
    
\end{itemize}

The partial derivative $\partial H_{t}/\partial \psi$ evaluated at $\psi = 0$ is:

\begin{equation} \label{eq_preisach_characteristic_dpsi}
    \left. \frac{\partial H_{t}\{i(t, \psi)\}}{\partial \psi} \right|_{\psi = 0} = \partial c\{i_{B}(t), w(t), \rho_{0}, \sigma_{0}\} + \, \epsilon\{\rho_{0}\}\partial h\{\rho_{0}, \sigma_{0}\} + 2 \!\!\! \sum_{j = 1}^{N(t, 0)-1} \epsilon\{\rho_{j}\}\, \partial h\{\rho_{j}, \sigma_{j}\}
\end{equation}

This completes the evaluation of $\partial H_{t}/\partial \psi$. The next step in forming the frequency coupling matrix is to evaluate the integrals of (\ref{eq_dlambda_coeff_real}) and (\ref{eq_dlambda_coeff_imag}). Given, the complexity of (\ref{eq_preisach_characteristic_dpsi}), numerical integration is more appropriate than symbolic integration. Lastly, the formation of $P^{(1)}$ and $P^{(2)}$ matrices from $\partial\Lambda_{n}^{\Re}/\partial \psi$ and $\partial\Lambda_{n}^{\Im}/\partial \psi$ is analogous to the formation of $Y^{(1)}$ and $Y^{(2)}$ in Section \ref{sec_general_harmonic_domain}.

\section{Results}
\label{sec_results_main}

\begin{figure}
    \centering
    \includegraphics[scale=1]{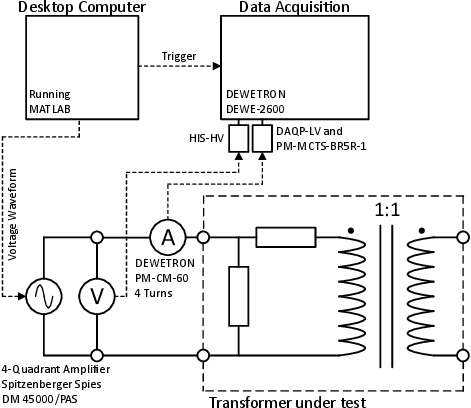}
    \caption{Test configuration for small transformer.}
    \label{fig_test_setup}
\end{figure}

Two series of tests are performed on a small transformer. The first series provides major loop data to implement the time-periodic Preisach model. The second series applies voltage perturbations to the transformer's primary side winding to construct a frequency coupling matrix and verify theoretical results. Both series of tests have the same setup shown in Fig. \ref{fig_test_setup}. 

The transformer under test is a 230 V single-phase unit with primary and secondary windings with the same number of turns. The primary side winding is defined as the side of applied voltage, and the secondary side winding is in open circuit. A test voltage is generated from a linear amplifier with a prescribed waveform sent from a computer. Terminal voltage and current are both sampled at 100 kHz. Low-pass filtering is applied to both current and voltage signals to remove transducer artifacts. The measured voltage is numerically integrated to obtain the equivalent flux linkage, which also includes the effects of eddy current and conduction losses. The equivalent flux linkage and open-circuit current are the two waveforms used to create the time-periodic Preisach model. 

\subsection{Implementing the Time-Periodic Preisach Model}
\label{sec_test_preisach}

This section explains how the rising and dropping equivalent flux linkages, $\lambda_{r}(i, \gamma_{m})$ and $\lambda_{d}(i, \gamma_{m})$, are formulated from test results using cubic splines. The shaping function, $h(\beta_{m}, \alpha_{m})$, is constructed according to the approximate method of Section \ref{sec_approximate_method}. It is shown that the internal minor loops condition, $m_{0}(\gamma_{m}) \geq 0$, cannot be held for the tested transformer. Therefore, the generating function method of Section \ref{sec_generating_method} is not used, as it is numerically disadvantaged to the approximate method. Lastly, the common mode function is implemented according to Section \ref{sec_common_mode} to complete the implementation of the time-periodic Preisach model.

\subsubsection{Representing $\lambda_{r}(i, \gamma_{m})$ and $\lambda_{d}(i, \gamma_{m})$ with Cubic Splines}

Rising and dropping equivalent flux linkages are represented by two-dimensional cubic splines $\tilde{\lambda}_{r}$ and $\tilde{\lambda}_{d}$:

\begin{equation} \label{eq_spline_r_def}
    \lambda_{r}(i, \gamma_{m}) = \tilde{\lambda}_{r}(\zeta, \gamma_{m})
\end{equation}
\begin{equation} \label{eq_spline_d_def}
    \lambda_{d}(i, \gamma_{m}) = \tilde{\lambda}_{d}(\zeta, \gamma_{m})
\end{equation}

\noindent where $\zeta = i / \gamma_{m}$. The domain of $\zeta$ is the interval $[-1, 1]$ because $-\gamma_{m} \leq i \leq \gamma_{m}$, which is divided into 300 smaller intervals in the cubic spline. The boundary points between the smaller intervals are $\zeta_{0}=-1$, $\zeta_{1}$, $\ldots$, $\zeta_{k}$, $\ldots \zeta_{300}=1$. For the maximum current, $\gamma_{m} \in [0, \gamma_{M}]$, its domain is split into 12 smaller intervals, where the interval boundaries are indexed by $l$ to give $\gamma_{m,l}$. Equations (\ref{eq_spline_r_def}) and (\ref{eq_spline_d_def}) are not well defined at $i=0$ and $\gamma=0$. To alleviate this issue, the origin is specially defined as $\lambda_{r}(0, 0) = \tilde{\lambda}_{r}(0, 0)$ and similarly for $\lambda_{d}$. The two-dimensional cubic spline for $\tilde{\lambda}_{r}$, and similarly for $\tilde{\lambda}_{d}$, in the interval $\zeta \in [\zeta_{k}, \zeta_{k+1}]$ and $\gamma_{m} \in [\gamma_{m,l}, \gamma_{m,l+1}]$  is defined by the polynomial equation:

\begin{equation} \label{eq_cubic_spline_structure}
    \tilde{\lambda}_{r}(\zeta, \gamma_{m}) = \sum_{g = 0}^{3} \sum_{h = 0}^{3} A_{g,h,k,l}(\zeta - \zeta_{k})^{g}(\gamma_{m} - \gamma_{m,l})^{h}
\end{equation}

\noindent where $A_{g,h,k,l}$ are polynomial coefficients. Cubic splines have the property that all first and second order partial derivatives are continuous including at the boundaries $\zeta_{k}$ and $\gamma_{m,l}$. Therefore, the two-dimensional cubic splines inherently satisfy Condition 2 of Section \ref{sec_permissible_data}. The main advantage for employing cubic splines is they have well defined derivatives, which is needed for forming frequency coupling matrix from Section \ref{sec_preisach_harmonic_domain}. The process for determining the polynomial coefficients is explained. 

\subsubsection{Forming $\tilde{\lambda}_{r}$ and $\tilde{\lambda}_{d}$}
\label{sec_data_cleaning}

This section explains how the test data for current $i_{l}(t)$ and equivalent flux linkage $\lambda_{l}(t)$ are converted to a cubic spline. The index $l$ is a reference to the test number. The first test, $l=1$, applied a 20 Vrms voltage to the test transformer. A further 11 tests were conducted, increasing the voltage by 20 Vrms each time and stopping at the maximum of 240 Vrms. For each test, the open-circuit current would go through a transient phase, once this had settled, the voltage and current were recorded for 240 ms. Then the voltage time series was integrated to give the equivalent flux linkage. Linear bias is removed from the current and equivalent linkage flux time series so that their minimum and maximum values align over the 12 cycles of the recording. Beginning at the time of the first current minimum, each time series is averaged over the next 11 cycles. The results of this process are $i_{l}(t)$ and $\lambda_{l}(t)$. Time has a sampling rate of $T_{s} = $ 10 $\mu$s so that $i_{l}(t)$ and $\lambda_{l}(t)$ is known at times $t = n T_{s}$, where $n \in \mathbb{Z}$.

It is unlikely that $\min\{i_{l}(t)\} = -\max\{i_{l}(t)\}$ will be exactly true at this step in the data processing. This condition is necessary for the definitions (\ref{eq_reduced_r_def}) and (\ref{eq_reduced_d_def}) to be valid and for applying the techniques of Section \ref{sec_fitting_reduced}. Therefore, a near linear transformation $f_{l}(i)$ is applied to $i_{l}(t)$ to satisfy this requirement:

\begin{equation}
    f_{l}(i) = a_{l}i^2 + b_{l}i + c_{l} \approx i + c_{l}
\end{equation}

\noindent The polynomial coefficients $a_{l}$, $b_{l}$ and $c_{l}$ are chosen so that $\min\{f_{l}(i_{l}(t))\} = -\max\{f_{l}(i_{l}(t))\}$ and for the dc component of $f_{l}(i_{l}(t))$ to be zero. This step introduces a small error to the model.

\begin{figure}
    \centering
    \includegraphics[scale=0.38]{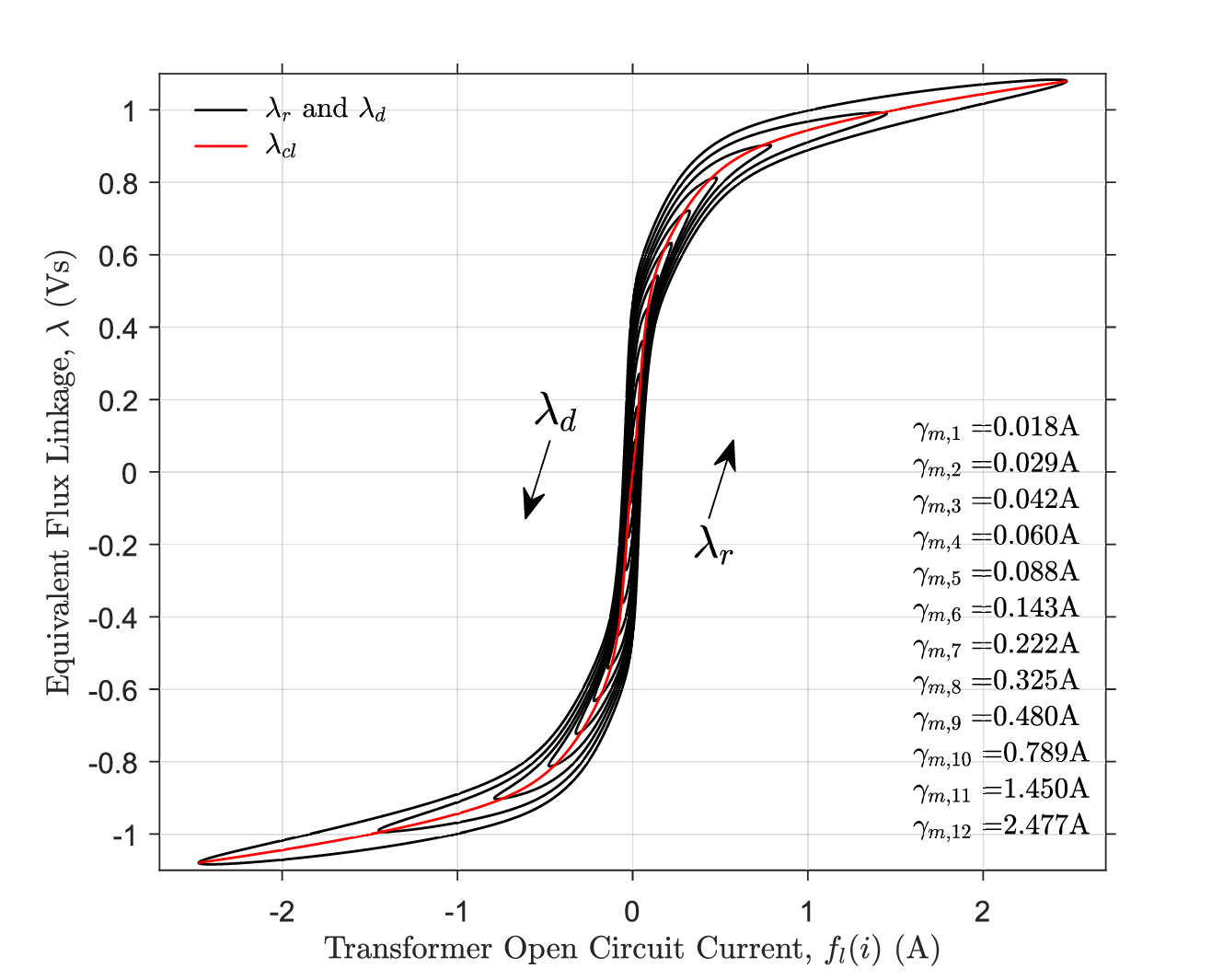}
    \caption{Hysteresis loops of the test transformer compared with its centre line model.}
    \label{fig_hysteresis_loop_all}
\end{figure}

A numerical relationship is developed between equivalent flux linkage and the modified open-circuit test current $f_{l}(i_{l}(t))$. Current has a minimum at $t = t_{\text{min},l} = 0$ and $t = T$. Also, a maximum current occurs at a time $t_{\text{max},l}$, which separates the period where $f_{l}(i_{l}(t))$ is increasing and decreasing. For $t \geq t_{\text{min},l}$ and $t \leq t_{\text{max},l}$, the open-circuit current is monotonically increasing, which defines the rising current equivalent flux linkage curve for $\gamma_{m} = \gamma_{m,l} = \max\{f_{l}(i_{l}(t))\}$:

\begin{equation}
    \lambda_{r,l}(i) = \lambda_{l}(t) \quad \text{where} \quad i = f_{l}(i_{l}(t))
\end{equation}

\noindent for a unique time $t$ in the interval $[t_{\text{min},l}, t_{\text{max},l}]$. Similarly, the decreasing current equivalent flux linkage is:

\begin{equation}
    \lambda_{d,l}(i) = \lambda_{l}(t) \quad \text{where} \quad i = f_{l}(i_{l}(t))
\end{equation}

\noindent for a unique time $t$ in the interval $[t_{\text{max},l}, t_{\text{min},l}+T]$. The test results for the transformer are summarised in the hysteresis loops of Fig. \ref{fig_hysteresis_loop_all}, which demonstrates a strong symmetry where $\lambda_{r,l}(i) \approx -\lambda_{d,l}(-i)$.

The cubic splines $\tilde{\lambda}_{r}$ and $\tilde{\lambda}_{d}$ optimally fit $\lambda_{r,l}$ and $\lambda_{d,l}$, respectively. The optimisation process consists of two steps: the first step is an optimal selection of the coefficients $A_{g,0,k,l}$ for each test $l$; the second step optimally selects the coefficients $A_{g,h,k,l}$ for $h \geq 1$. The objective of the first step is minimise the error for each $l$:

\begin{equation} \label{eq_opt_lambda_r_step1}
    \min_{A_{g,0,k,l}} \, \sum_{i_{n}\geq -\gamma_{m,l}}^{i_{n} \leq \gamma_{m,l}} \big(\tilde{\lambda}_{r}(i_{n}/\gamma_{m,l}, \gamma_{m,l}) - \lambda_{r,l}(i_{n}) \big)^{2}
\end{equation}

\noindent where $i_{n}=f_{l}(i_{l}(nT_{s}))$ for $t_{\text{min},l} \leq nT_{s} \leq t_{\text{max},l}$. The optimisation is subject to conditions upon $A_{g,0,k,l}$ to ensure the continuity of $\partial^{2}\tilde{\lambda}_{r}/\partial \zeta^{2}$ at the interval boundaries $\zeta_{k}$. Furthermore, terminal conditions at $\zeta = -1$ and $\zeta = 1$ require $\tilde{\lambda}_{r}(-1, \gamma_{m,l}) = \lambda_{r,l}(-\gamma_{m,l})$ and $\tilde{\lambda}_{r}(1, \gamma_{m,l}) = \lambda_{r,l}(\gamma_{m,l})$. The terminal conditions ensure (\ref{eq_loop_bottom}) and (\ref{eq_loop_top}) are satisfied, as the optimisation formulation for $\tilde{\lambda}_{d}$ has the same requirement. The optimisation formulation for $\tilde{\lambda}_{d}$ is identical to $\tilde{\lambda}_{r}$ except with $t_{\text{max},l} \leq nT_{s} \leq t_{\text{min},l}+T$. The optimisation (\ref{eq_opt_lambda_r_step1}) is a Quadratic Programming (QP) problem with $A_{g,0,k,l}$ as the variables.

The second step optimisation converts 13 one-dimensional cubic splines, $\tilde{\lambda}_{r}(\zeta, \gamma_{m,l})$ for $l = 0, 1, ... 12$, into a single two-dimensional cubic spline $\tilde{\lambda}_{r}(\zeta, \gamma_{m})$. Note, the $l = 0$ result occurs when $\gamma_{m,l} = 0$, which is not derived from a test on the transformer. Rather, $A_{g,0,k,0} = 0$ under the assumption that no current is drawn from the test transformer when no voltage is applied. The objective is to minimise:

\begin{equation} \label{eq_opt_lambda_r_step2}
    \min_{\substack{A_{g,h,k,l} \\ h \geq 1}} \sum_{l = 1}^{12} \int_{\gamma_{m,l-1}}^{\gamma_{m,l}} \Bigg(\tilde{\lambda}_{r}(\zeta, \gamma) - \frac{\gamma_{m,l} - \gamma}{\Delta\gamma_{m,l}}\tilde{\lambda}_{r}(\zeta, \gamma_{m,l-1}) - \frac{\gamma - \gamma_{m,l-1}}{\Delta\gamma_{m,l}}\tilde{\lambda}_{r}(\zeta, \gamma_{m,l}) \Bigg)^{2} \,d\gamma
\end{equation}

\noindent where $\Delta\gamma_{m,l} = \gamma_{m,l} - \gamma_{m,l-1}$. This is to minimise the difference between $\tilde{\lambda}_{r}(1, \gamma_{m})$ and its linear interpolation between known points of $\gamma_{m} = \gamma_{m,l}$. This optimisation is also subject to constraints requiring the second order partial derivatives to be continuous. Solving (\ref{eq_opt_lambda_r_step2}) is achieved by formulation into a QP problem.

\subsubsection{Achievability of the Internal Minor Loops Condition}

This section assess whether the internal minor loops condition is achievable for the test transformer. The evaluation of $m_{0}(\gamma_{m})$ from Section \ref{sec_internal_minor} begins by differentiating $\lambda_{s}$ as required according to its construction in (\ref{eq_reduced_diff_conversion}), (\ref{eq_spline_r_def}), (\ref{eq_spline_d_def}) and (\ref{eq_cubic_spline_structure}). The derivative $\partial^{2}\lambda_{s}/\partial i \partial \gamma_{m}$ is evaluated at uniformly sampled values of $\gamma_{m}$, which creates a one-dimensional quadratic spline for each value. The absolute function  $|\partial^{2}\lambda_{s}/\partial i \partial \gamma_{m}|$ is created by including the zero crossings of $\partial^{2}\lambda_{s}/\partial i \partial \gamma_{m}$ as knots in the spline and swapping the sign of the quadratic function wherever $\partial^{2}\lambda_{s}/\partial i \partial \gamma_{m} < 0$. The integral term of (\ref{eq_g_requirement_1_short}) can be analytically calculated because $|\partial^{2}\lambda_{s}/\partial i \partial \gamma_{m}|$ is piecewise quadratic. The last step is evaluating $\partial \lambda_{se}/\partial i$ along the boundary line $(\gamma_{m}, \gamma_{m})$.

\begin{figure}
    \centering
    \includegraphics[scale=0.38]{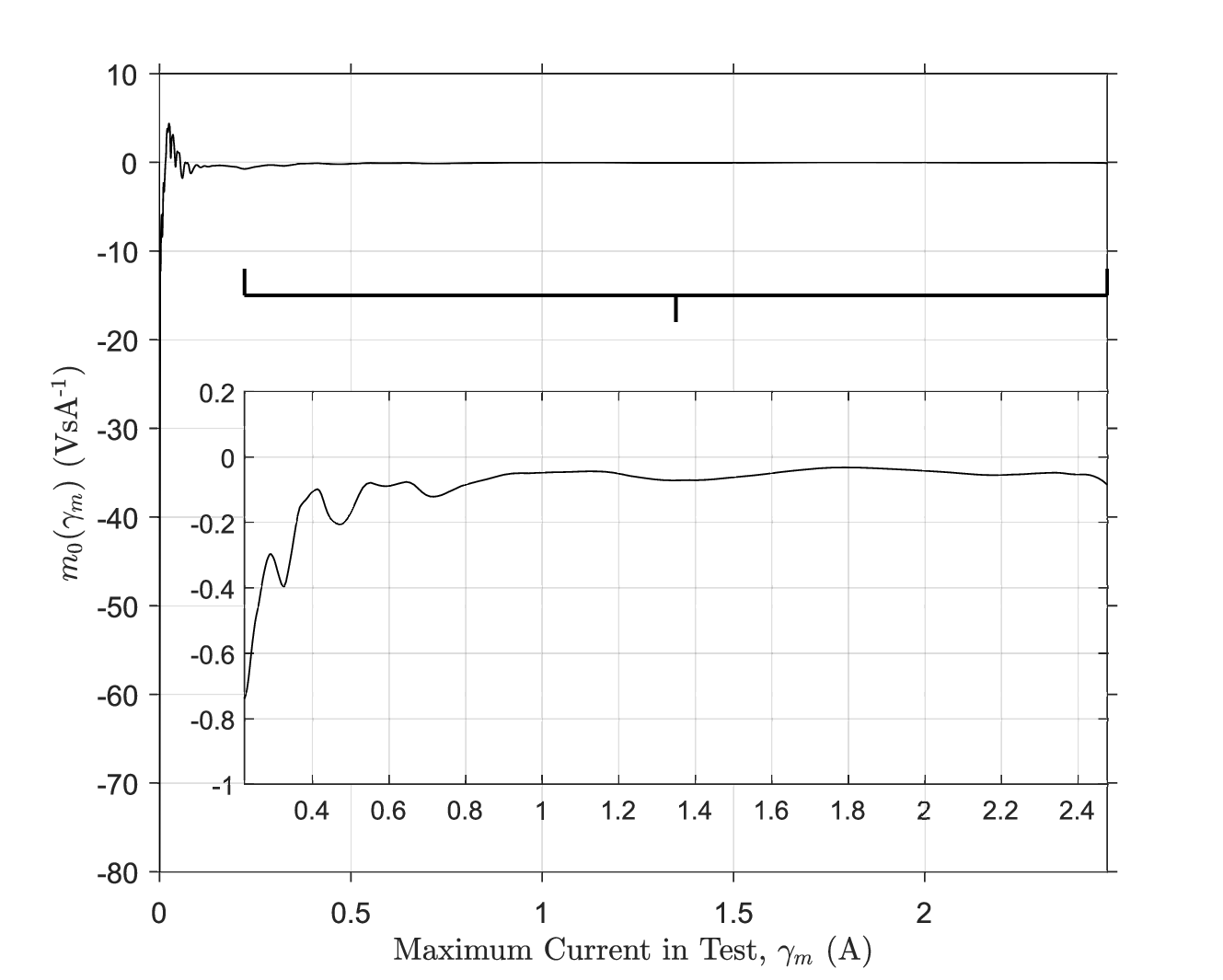}
    \caption{Test transformer's $m_{0}$ measure.}
    \label{fig_m1_measure}
\end{figure}

Fig. \ref{fig_m1_measure} shows that $m_{0}(\gamma_{m}) \leq 0$ for most $\gamma_{m}$; therefore, it is not always possible for $\mu(\beta, \alpha) \geq 0$. The following section uses the approximate method of Section \ref{sec_approximate_method} to construct the shape function $h(\beta_{m}, \alpha_{m})$.

\subsubsection{Approximate Shape Function}

\begin{figure}
    \centering
    \includegraphics[scale=0.34]{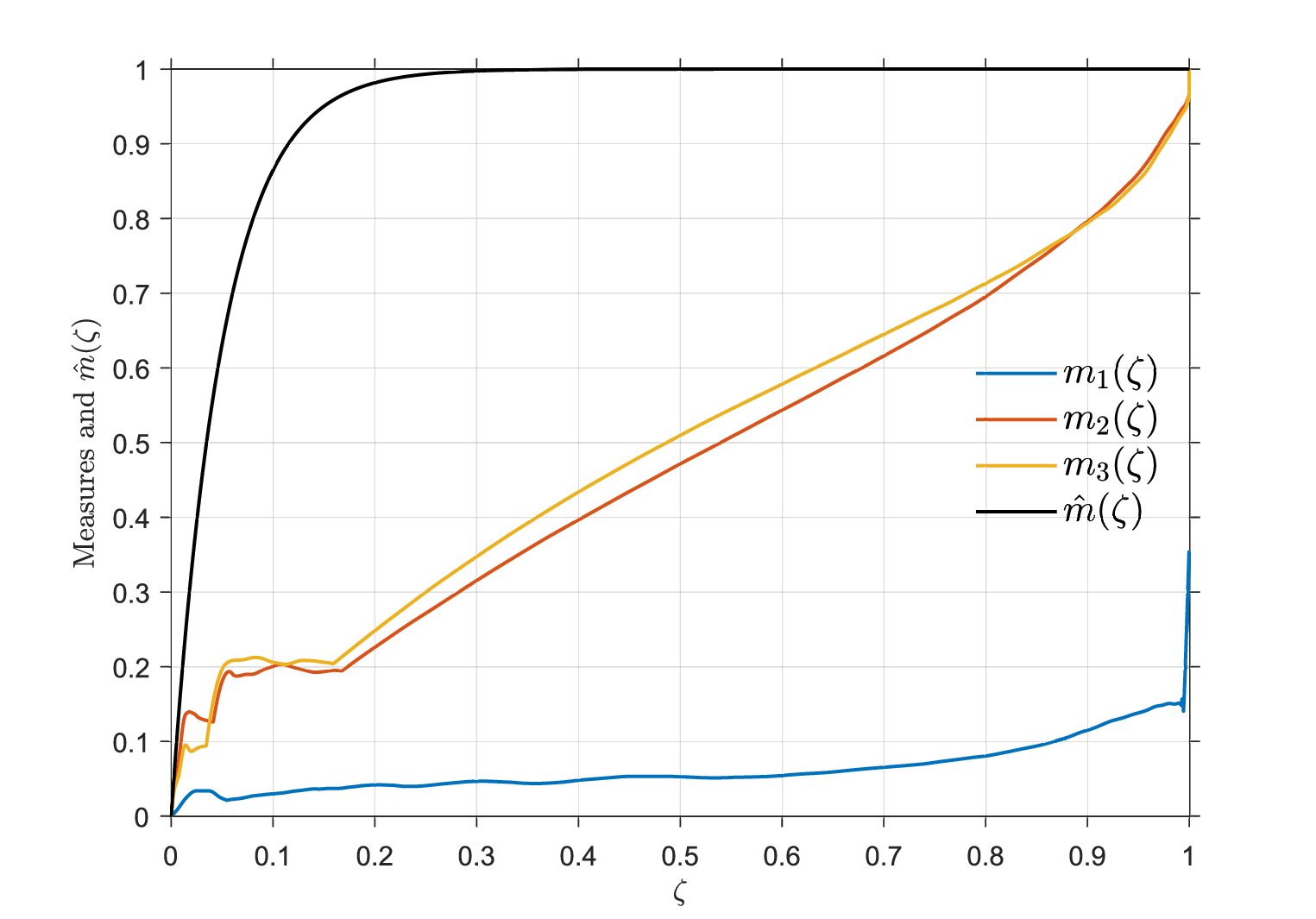}
    \caption{Setting test transformer's $\hat{m}(\zeta)$ function with $a$=20.}
    \label{fig_m123_measure}
\end{figure}

The approximate method selects a parameter $a$ in (\ref{eq_mhat_choice}) so that  $\hat{m}(\zeta) \geq m_{j}(\zeta)$ for all measures $j \in \{1, 2, 3\}$. Each measure is calculated according to (\ref{eq_m1_measure})-(\ref{eq_m3_measure}), where $\zeta$ and $\gamma$ are appropriately sampled. For the test transformer, Fig. \ref{fig_m123_measure} shows a value of at least $a\geq 20$ is necessary for $\hat{m}(\zeta) \geq m_{j}(\zeta)$, where the region of $\zeta < 0.05$ is most restrictive on $a$.

\begin{figure}
    \centering
    \includegraphics[scale=0.5]{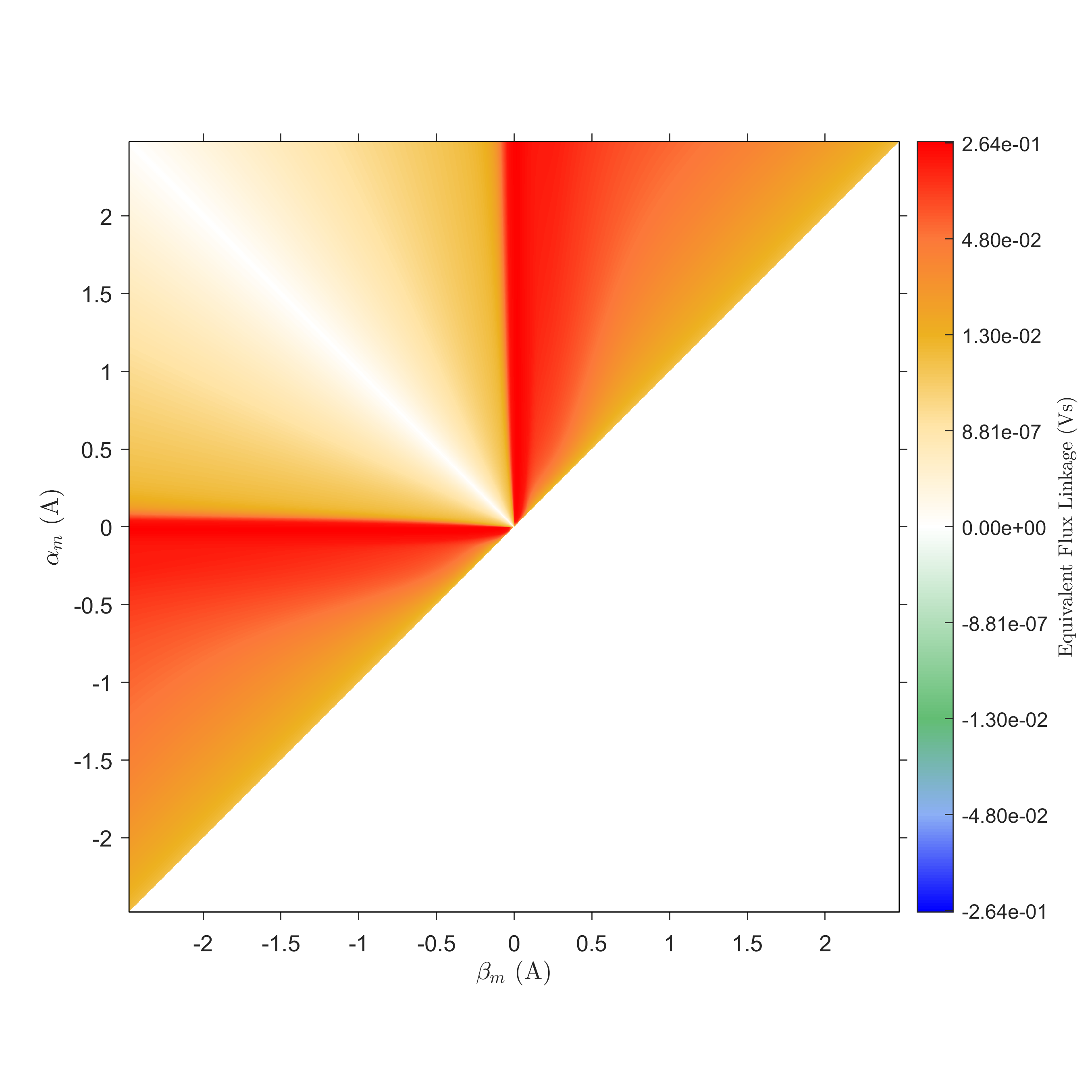}
    \caption{Shape function $h(\beta_{m}, \alpha_{m})$ for the tested transformer.}
    \label{fig_h_shaping_function}
\end{figure}

\begin{figure}
    \centering
    \includegraphics[scale=0.5]{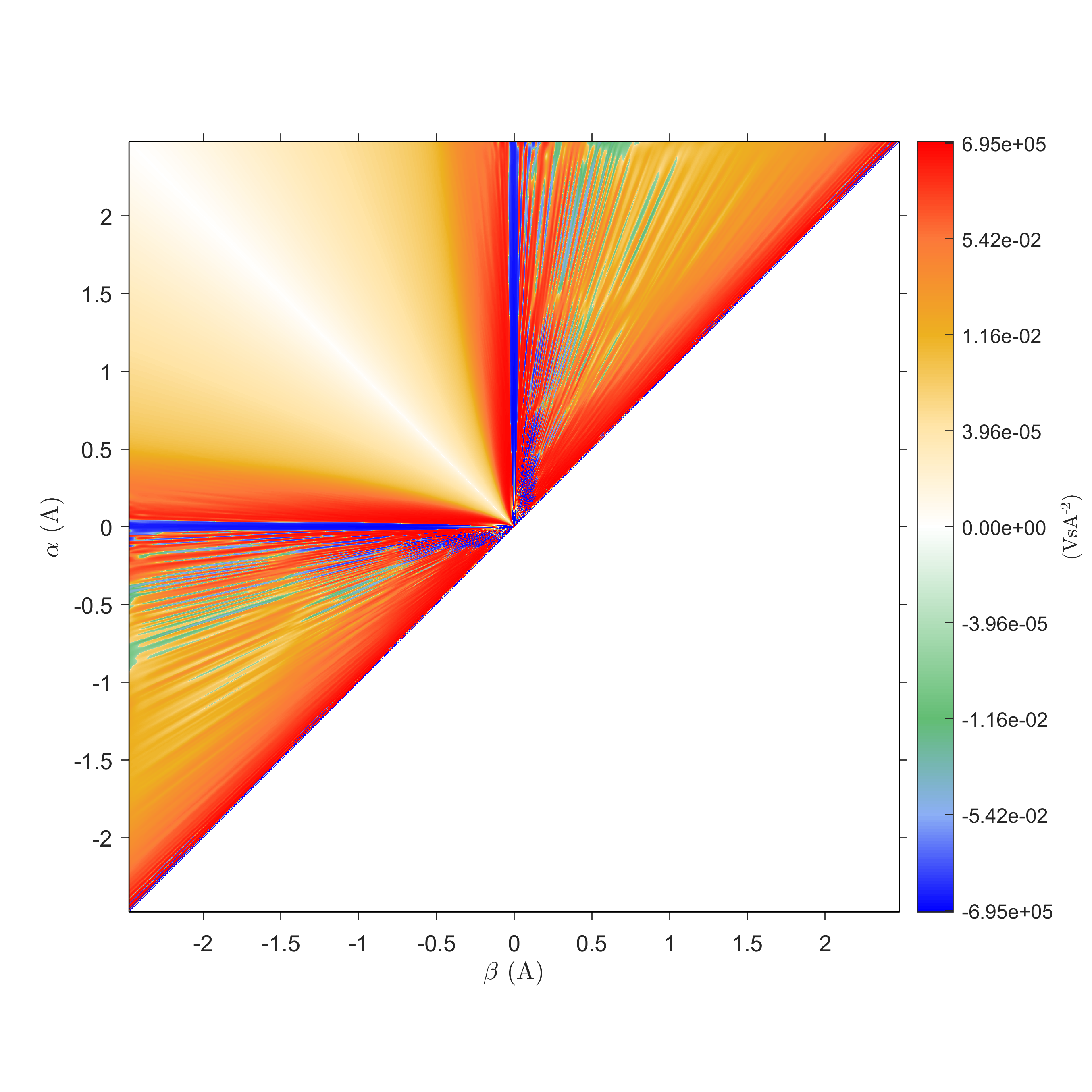}
    \caption{Preisach weight function $\mu(\beta, \alpha)$ for the tested transformer.}
    \label{fig_mu_function}
\end{figure}

Fig. \ref{fig_h_shaping_function} shows the results of calculating the shape function in the region $\beta_{m} \leq \alpha_{m}$. It is calculated according to (\ref{eq_h_explicit_approximate}) and (\ref{eq_m_half}) with the selected $\hat{m}(\zeta)$ from Fig. \ref{fig_m123_measure}. The distributed Preisach weight function is derived from the shaping function according to (\ref{eq_general_preisach_weight_formulas}) and presented in Fig. \ref{fig_mu_function}. The largely positive $\mu(\beta, \alpha) \geq 0$ indicates that the approximate method is successful. Internal minor loops can be expected in many instances. However, the large negative region around $\beta = 0$ and $\alpha = 0$ implies internal minor loops are not always possible around $i = 0$. The conditions for external minor loops can be inferred from (\ref{eq_internal_step2}) and Fig. \ref{fig_internal_construction}. Vertices of rectangular regions of $\mu(\beta, \alpha)$, which are overall negative, determine the parameters for $\beta_{m}$, $\beta_{n}$, $i$, $\alpha_{n}$ and $\alpha_{m}$, to cause external minor loops.  

\subsubsection{Common Mode Function}

\begin{figure}
    \centering
    \includegraphics[scale=0.5]{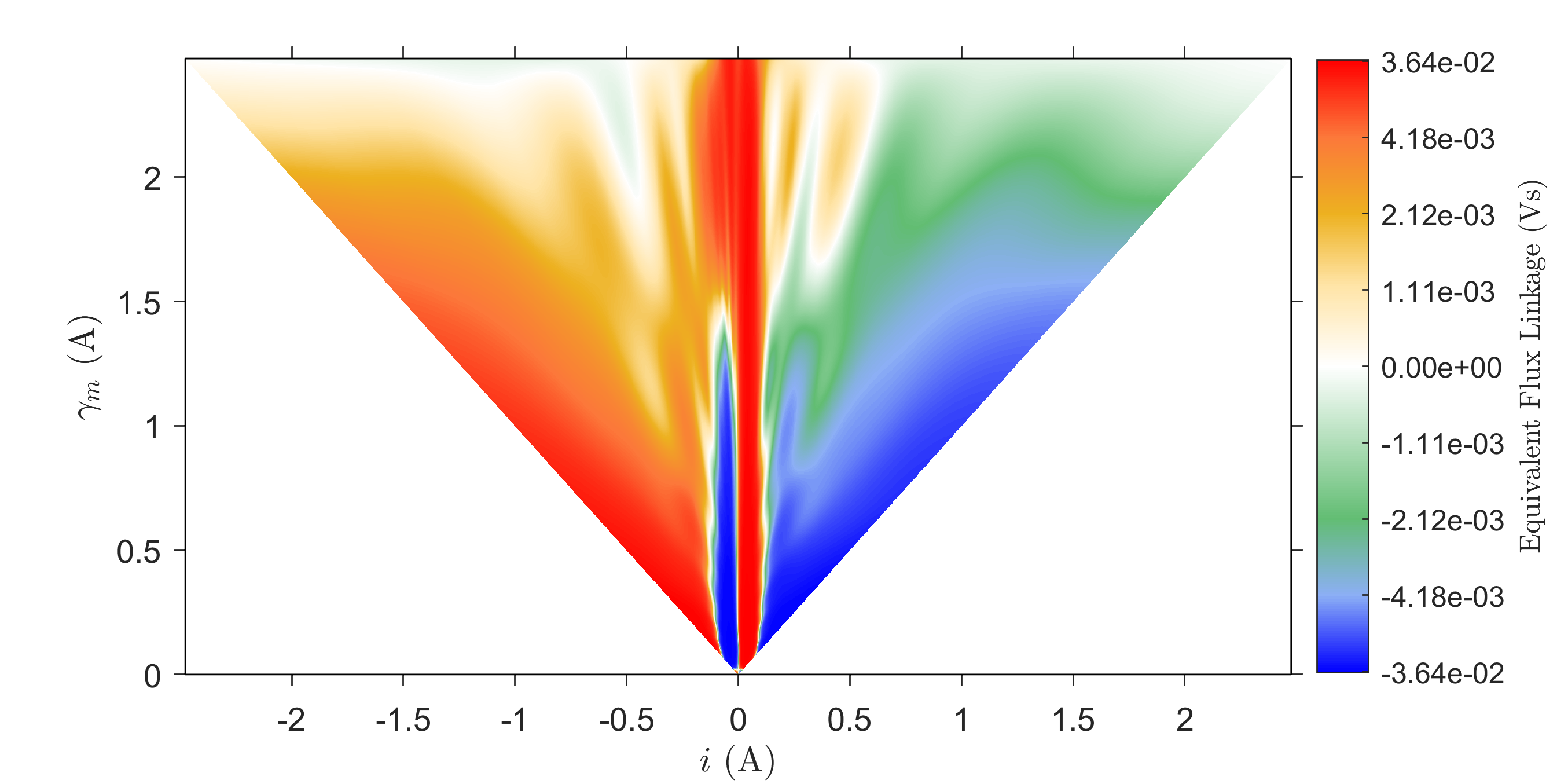}
    \caption{Common mode perturbation $\lambda_{\Delta c}(i, \gamma_{m})$ for the tested transformer.}
    \label{fig_lambda_c_deviation}
\end{figure}

The time-periodic Preisach model for the tested transformer is completed with the common mode function $c(i, \beta_{m}, \alpha_{m})$. Eq. (\ref{eq_reduced_common_conversion}) calculates the common mode component with $\lambda_{r}$ and $\lambda_{d}$ formed by two-dimensional cubic splines in (\ref{eq_spline_r_def}) and (\ref{eq_spline_d_def}). The centre line function $\lambda_{cl}(i)$ from (\ref{eq_center_line_formula}) is a one-dimensional cubic spline with $\gamma_{M} = \gamma_{m,12}=2.477$~A. Fig. \ref{fig_hysteresis_loop_all} demonstrates that the centre line function is the dominant relationship between the magnetising current and the equivalent flux linkage. Therefore, time-periodic Preisach models can reasonably have $\lambda_{\Delta c}(i, \gamma_{m}) = 0$ in (\ref{eq_common_mode_function_impl}). Nonetheless, $\lambda_{\Delta c}(i, \gamma_{m})$ is implemented with the result of its calculation shown in Fig. \ref{fig_lambda_c_deviation}.

\subsection{Time Domain Demonstration}

The improved accuracy of the time-periodic Preisach model is proved by comparison with the centre line model $\lambda_{cl}(i)$. Section \ref{sec_frequency_coupling_matrix} demonstrates improvement in the frequency domain while this section briefly considers results in the time domain. Two periodic voltages are applied to the test transformer according to the setup in Fig. \ref{fig_test_setup}. The recorded voltage and current measurements are cleaned following the process of Section \ref{sec_data_cleaning} to obtain $i_{l}(t)$ and $\lambda_{l}(t)$; the indices $l=13$ and $l=14$ are used for these tests. The time-periodic Preisach model (\ref{eq_preisach_characteristic}) is applied to $i_{l}(t)$ to determine how well it replicates $\lambda_{l}(t)$. Also, $\lambda_{cl}(i)$ is applied to $i_{l}(t)$.

\begin{figure}
    \centering
    \includegraphics[scale=0.36]{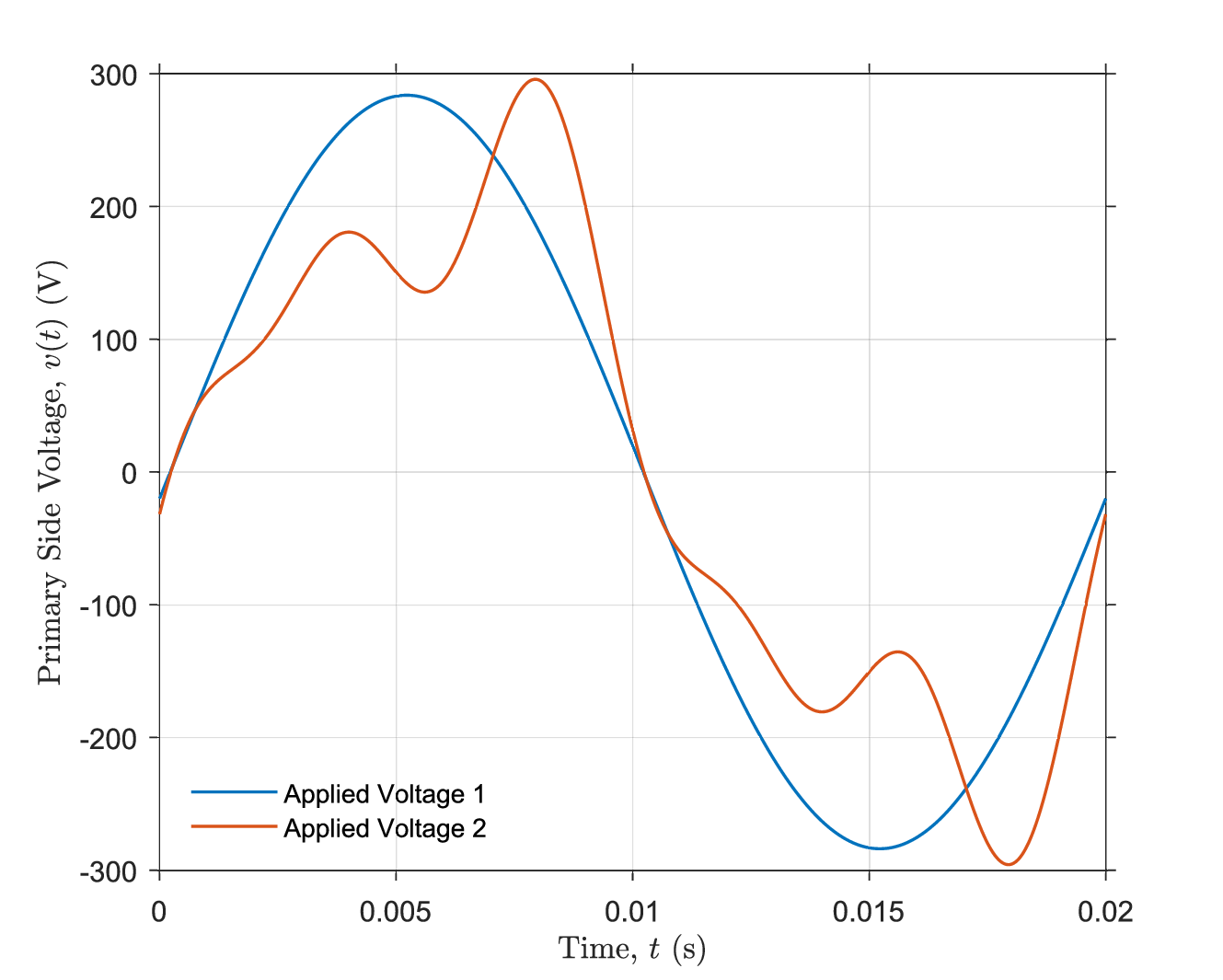}
    \caption{Time domain test voltages.}
    \label{fig_voltage_input_tests}
\end{figure}

\begin{figure}
    \centering
    \includegraphics[scale=0.36]{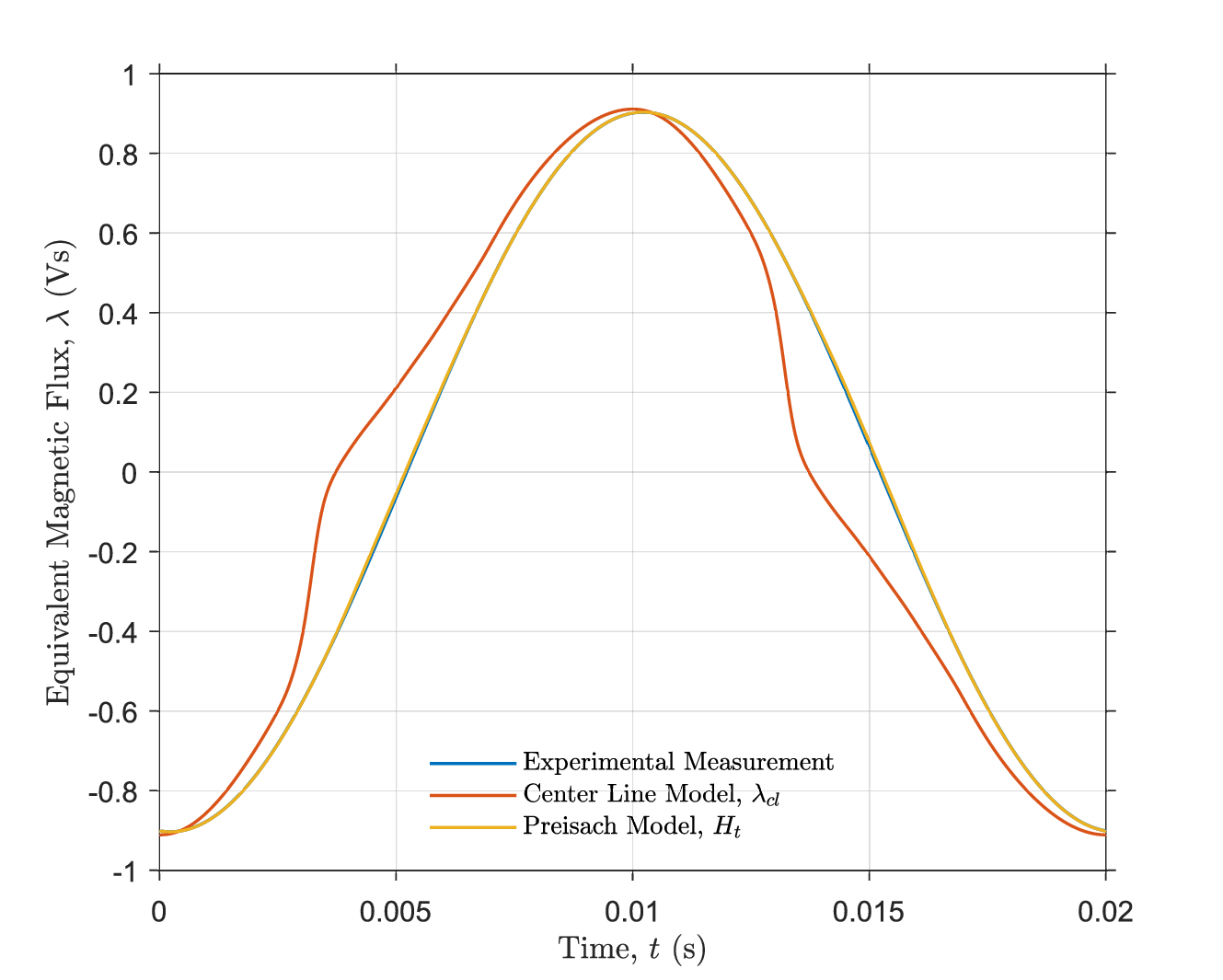}
    \caption{Model comparison of equivalent flux linkage for trained input.}
    \label{fig_test1_flux_linkage}
\end{figure}

Two test voltages are shown in Fig. \ref{fig_voltage_input_tests}. The first voltage replicates an early test for $\gamma_{m,10}$ with 200 Vrms at the primary side terminals. The time-periodic Preisach model is expected to closely replicate $\lambda_{10}(t)$ in $\lambda_{13}(t)$ as shown by the near overlap of curves in Fig. \ref{fig_test1_flux_linkage}. Small differences in results are because of slight changes in measurement between $l=10$ and $l=13$. Fig. \ref{fig_test1_flux_linkage} shows a significant improvement upon the centre line model. The largest error in the centre line model comes from currents when the height of the hysteresis loop is greatest, i.e. when current is close to zero.

\begin{figure}
    \centering
    \includegraphics[scale=0.36]{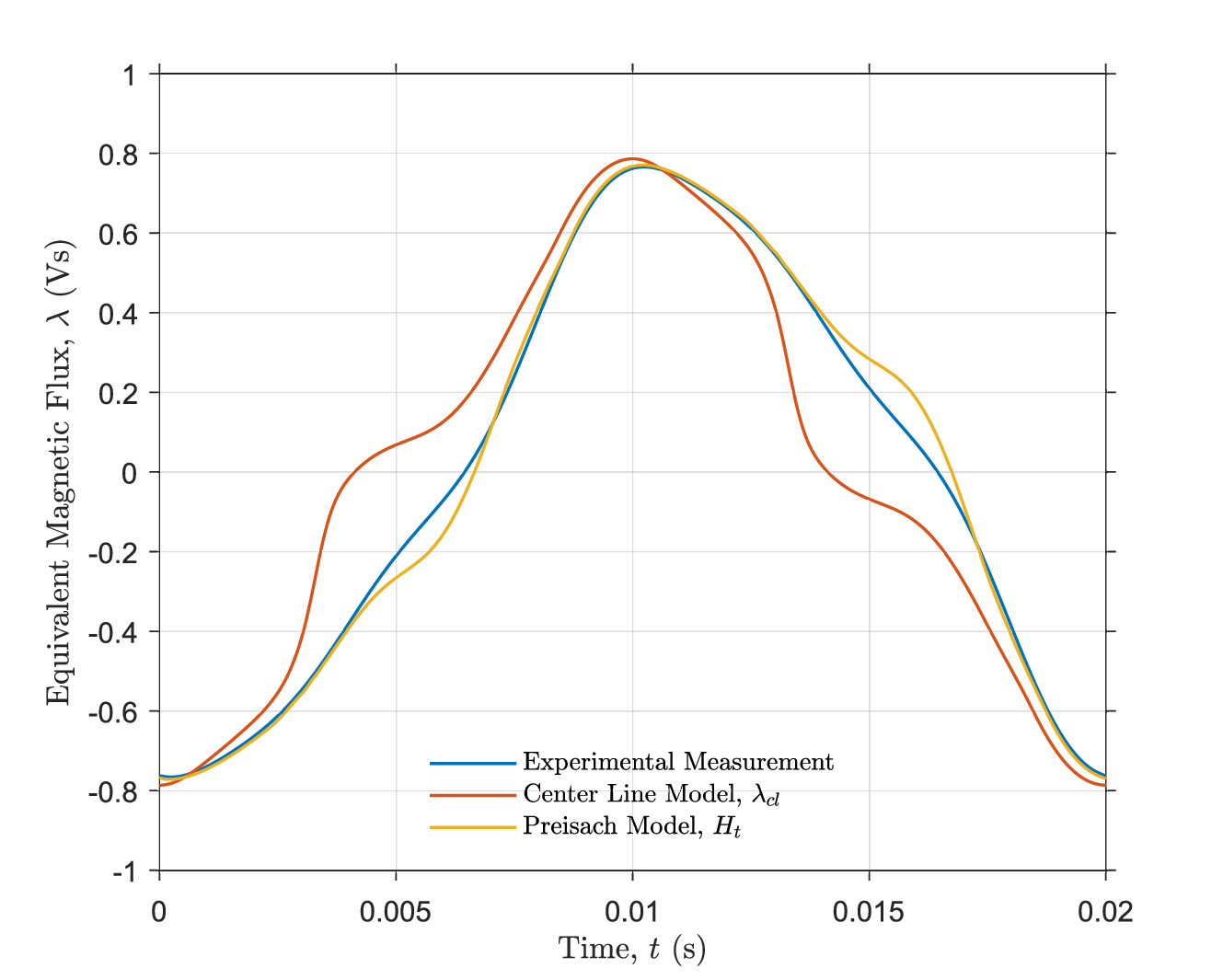}
    \caption{Comparison of equivalent flux linkage for untrained input.}
    \label{fig_test2_flux_linkage}
\end{figure}

The second voltage test applies a highly distorted waveform to assess accuracy to untrained data. The results of Fig. \ref{fig_test2_flux_linkage} demonstrates that the time-periodic Preisach model partially replicates experimental results. Future research could improve outcomes with more complex Preisach models, such as nonlinear, rate dependent, and dynamic versions. As expected, the centre line model again performs worse than the Preisach model.

\subsubsection{Asymmetric Current Demonstration}

\begin{figure}
    \centering
    \includegraphics[scale=0.36]{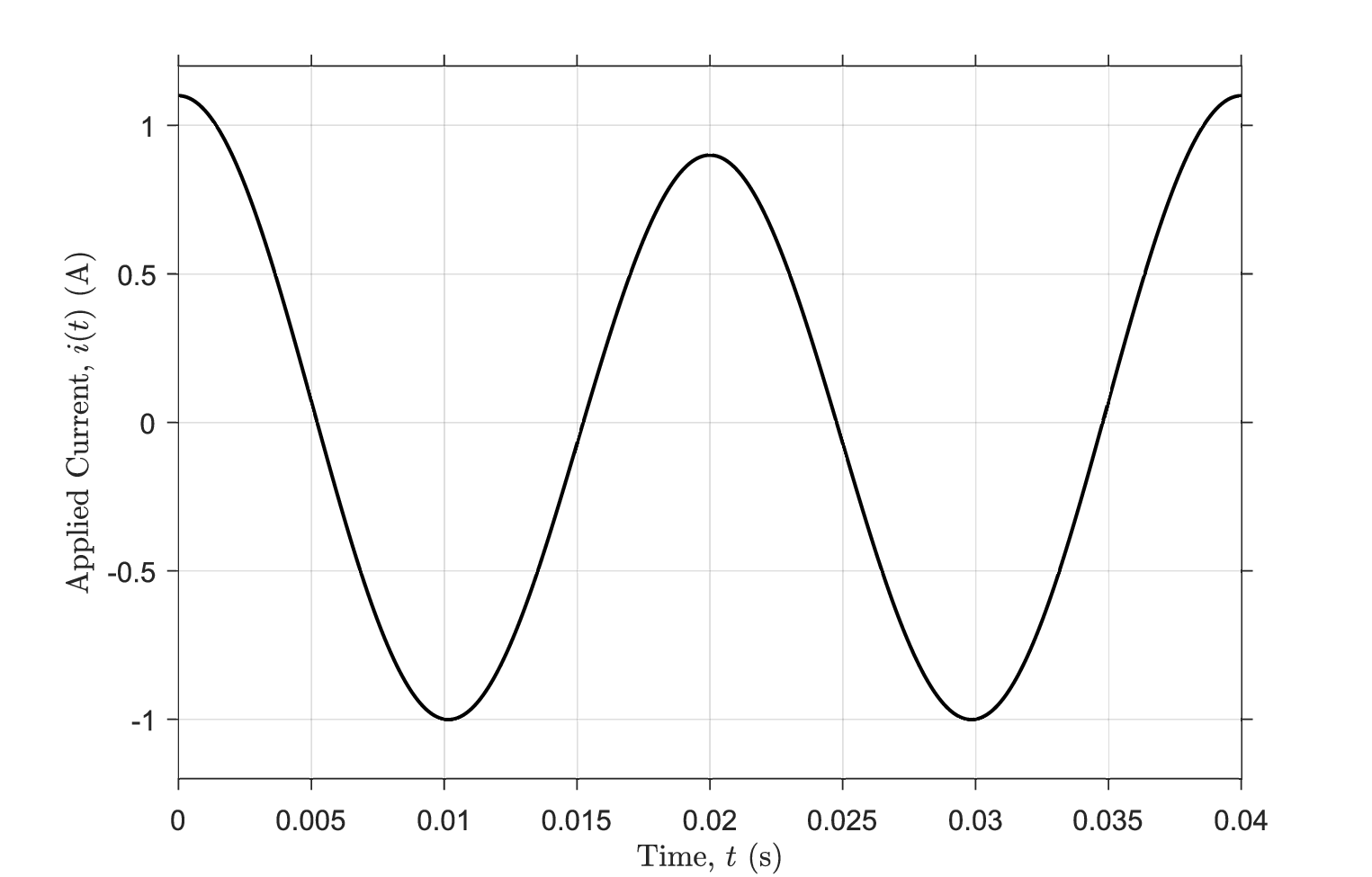}
    \caption{Asymmetric open-circuit current for testing the time-periodic Preisach model.}
    \label{fig_asymmetric_current_input}
\end{figure}

This section tests the time-periodic Preisach model with an asymmetric current to test the appropriateness of the splitting function $d(i, \gamma_{m})$. The previous section demonstrated accuracy for voltage waveforms composed entirely of odd harmonics. These tests create near symmetric curves on the $(i, \lambda)$ plane like in Fig. \ref{fig_hysteresis_loop_all}. Furthermore, they created open-circuit currents with one minimum and one maximum of equal magnitude but opposite sign, which results in the splitting function $d(i, \gamma_{m})$ being cancelled out in the implementation of (\ref{eq_preisach_characteristic}). Therefore, a theoretic asymmetric open-circuit current test is conducted with a 25 Hz sub-harmonic added to a fundamental 50 Hz component shown in Fig. \ref{fig_asymmetric_current_input}.

\begin{figure}
    \centering
    \includegraphics[scale=0.36]{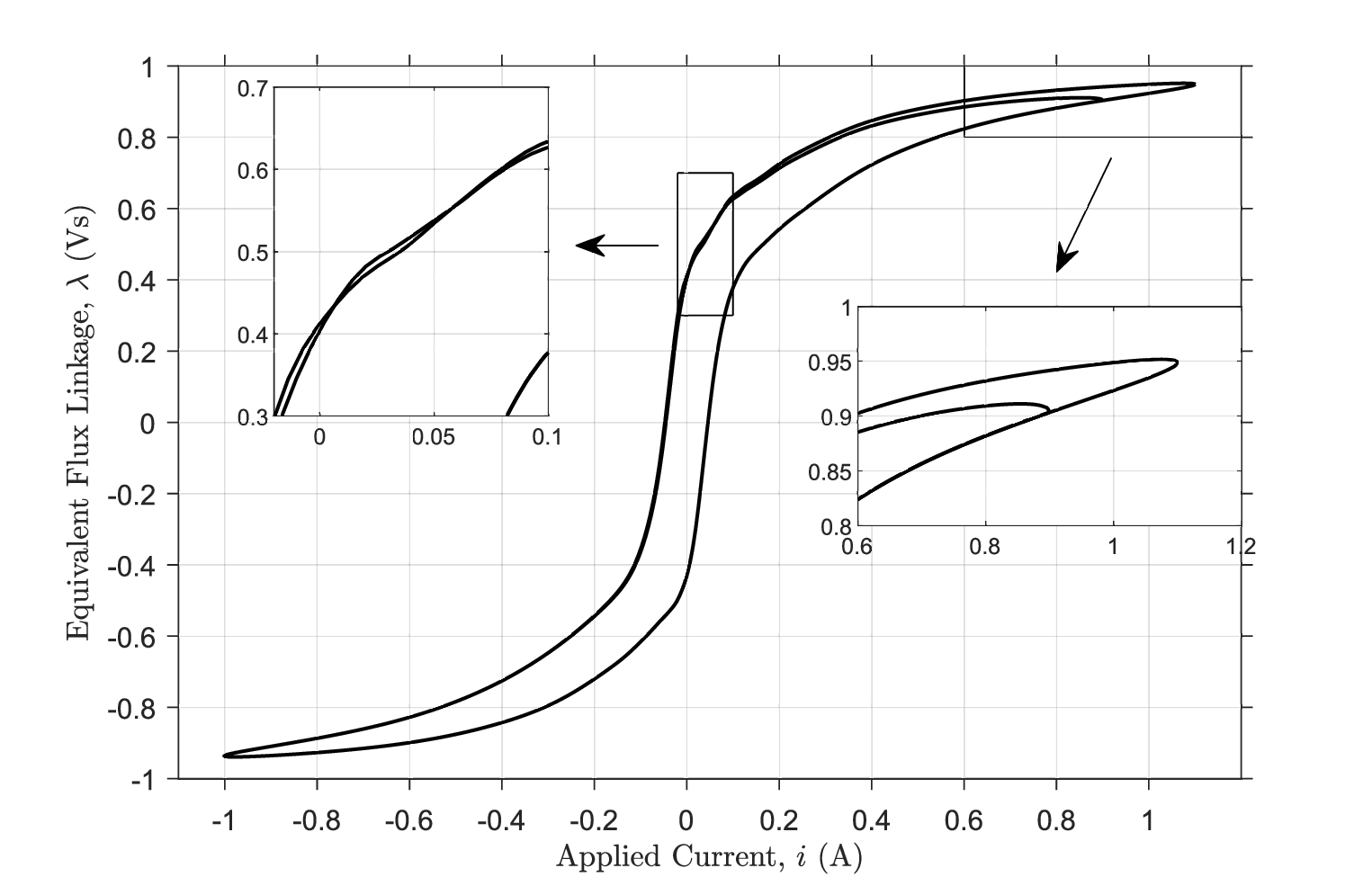}
    \caption{Hysteresis loop for asymmetric open-circuit current from Fig. \ref{fig_asymmetric_current_input}.}
    \label{fig_asymmetric_hysteresis_output}
\end{figure}

The results of the simulation are shown in the hysteresis loop of Fig. \ref{fig_asymmetric_hysteresis_output}. As expected, the lower peak current at 0.02 seconds gives a lower flux linkage than at 0.04 seconds, which is seen in the top right corner of Fig. \ref{fig_asymmetric_hysteresis_output}. Overall, the profile of the minor loop follows the major loop well, except for when the minor loop traces outside the major loop close to $i = 0$.  

\subsection{Frequency Coupling Matrix}
\label{sec_frequency_coupling_matrix}

This section demonstrates the improved accuracy of the time-periodic Preisach model in constructing the frequency coupling matrix over the centre line model. Section \ref{sec_preisach_harmonic_domain} provides the methodology of constructing the frequency coupling matrix for the time-periodic Preisach model. The frequency coupling matrix of the centre line model is calculated by omitting all but $\lambda_{cl}(i)$. Each model is compared against a benchmark frequency coupling matrix produced from a series of perturbations to the test transformer.

The frequency coupling matrix is constructed around an initialisation $i_{B}(t)=i_{13}(t-t_{s})$. This is when a 200 Vrms is applied to the test transformer with $v_{B}(t) = 200\sqrt{2}\cos(\omega t)$ and $\omega = 2\pi \times 50$ rads\textsuperscript{-1}. A time shift $t_{s}$ is added to $i_{13}(t)$ so that fundamental component $V_{B,1}$ of $v_{B}(t)$ lies along the real axis. All measured phases are relative to $\angle V_{B,1}$.

The construction of the benchmark frequency coupling matrix begins by applying multiple perturbed voltages $v_{B}(t) + \Delta v(t)$ to the transformer according to the test setup in Fig. \ref{fig_test_setup}. Each perturbation is of the form $\Delta v(t) = m \Delta v \sqrt{2} \cos(m\omega t + 2\phi \pi/N_{\phi})$. Harmonic order is increased from the fundamental component $m=1$ up to the $m=11$ harmonic. Perturbation voltage magnitude $m\Delta v$ is increased with harmonic order to keep the perturbation magnitude of the equivalent flux linkage constant, where $\Delta v = 2$ V. Also, for each harmonic order, phase is perturbed in twelve steps ($N_{\phi}=12$) from $\phi = 0$ to $\phi = 11$. 

The measured results of each test are cleaned as in Section \ref{sec_data_cleaning} while retaining the phase information relative to $V_{B,1}$. The harmonic components of each measured current are calculated from (\ref{eq_i_coeff_V1}) to give $I_{n,m,\phi}$. The calculation of the apparent admittance for each perturbation is the basis for forming $Y^{(1)}$ and $Y^{(2)}$. The apparent admittance for $I_{n}$ depending on $V_{m}$ with a phase shift $2\pi\phi/N_{\phi}$ is:

\begin{equation}
    Y_{n,m,\phi} = \frac{I_{n,m,\phi} - I_{B,n}}{m\Delta v \sqrt{2}e^{j2\pi\phi/N_{\phi}}}
\end{equation}

\noindent where $I_{B,n}$ are the harmonic coefficients of $i_{B}(t)$. The values of $Y^{(1)}_{n,m}$ and $Y^{(2)}_{n,m}$ can be inferred from $Y_{n,m,\phi}$, as these points should outline a circle. The average of the points $Y_{n,m,\phi}$ determines the centre of the circle:

\begin{equation}
    Y^{(1)}_{n,m} = \frac{1}{N_{\phi}}\sum_{\phi = 0}^{N_{\phi}-1} Y_{n,m,\phi}
\end{equation}

The average distance from the centre of the circle to $Y_{n,m,\phi}$ determines:

\begin{equation}
    |Y^{(2)}_{n,m}| = \frac{1}{N_{\phi}} \sum_{\phi = 0}^{N_{\phi}-1} |Y_{n,m,\phi} - Y^{(1)}_{n,m}|
\end{equation}

The method for calculating the phase angle of $Y^{(2)}_{n,m}$ can be inferred from $\Delta I_{n}/\Delta V_{m}=Y^{(1)}_{n,m}+Y^{(2)}_{n,m}(\overline{\Delta V_{m}}/\Delta V_{m})$, which gives:

\begin{equation}
    \angle Y^{(2)}_{n,m} = \frac{1}{N_{\phi}} \sum_{\phi = 0}^{N_{\phi}-1} \bigg( \Big( \angle \big(Y_{n,m,\phi} - Y^{(1)}_{n,m} \big) + 4\phi\pi/N_{\phi} \Big) \,\, \text{mod} \, 2\pi \bigg)
\end{equation}

\noindent where modular arithmetic is required and phase angles have to be carefully wrapped for the average to be meaningful. 

The result of forming the frequency coupling matrices are shown in Table \ref{tbl_frequency_coupling_matrix} up to the fifth harmonic. The time-periodic Preisach model improves upon the centre line model in the accuracy of $Y^{(1)}_{n,m}$ and $Y^{(1)}_{n,m}$ where $n$ and $m$ are odd integers. The largest improvement comes from reducing the error in the phase. For the $n=1$ and $m=1$ component of $Y^{(1)}$, the expected phase improves from -90\textdegree  to -87\textdegree 
 with -83\textdegree  being the target from the perturbation results.

\begin{table}[width=1.0\linewidth,cols=16,pos=h]
\caption{Frequency coupling admittance matrices for transformer open-circuit characteristics.}\label{tbl_frequency_coupling_matrix}
\begin{tabular*}{\tblwidth}{@{} |c|ccccc|ccccc|ccccc| @{} }
\toprule
& \multicolumn{5}{c|}{Centre Line Model, $\lambda_{cl}$} & \multicolumn{5}{c|}{Time-Periodic Preisach Model, $H_{t}$} & \multicolumn{5}{c|}{Perturbation Test Results} \\
$m$ & 1 & 2 & 3 & 4 & 5 & 1 & 2 & 3 & 4 & 5 & 1 & 2 & 3 & 4 & 5\\
\midrule
$n$ & \multicolumn{15}{c|}{$|Y^{(1)}|$ (mS)} \\
\midrule
1 & 5.23 & 0.00 & 1.15 & 0.00 & 0.30        & 5.09 & 0.24 & 1.10 & 0.18 & 0.26      & 5.07 & 0.02 & 1.14 & 0.01 & 0.27 \\
2 & 0.00 & 2.61 & 0.00 & 0.87 & 0.00        & 0.38 & 2.58 & 0.23 & 0.84 & 0.19      & 0.03 & 1.53 & 0.02 & 0.69 & 0.03 \\
3 & 3.46 & 0.00 & 1.74 & 0.00 & 0.69        & 3.42 & 0.22 & 1.73 & 0.19 & 0.68      & 3.41 & 0.03 & 1.79 & 0.02 & 0.74 \\
4 & 0.00 & 1.73 & 0.00 & 1.31 & 0.00        & 0.21 & 1.89 & 0.17 & 1.39 & 0.17      & 0.01 & 1.35 & 0.03 & 1.31 & 0.02 \\
5 & 1.49 & 0.00 & 1.15 & 0.00 & 1.05        & 1.42 & 0.15 & 1.19 & 0.14 & 1.07      & 1.41 & 0.02 & 1.22 & 0.02 & 1.14 \\
\midrule
$n$ & \multicolumn{15}{c|}{$\angle Y^{(1)}$ (Degrees)} \\
\midrule
1 & -90 & 0 & 77 & 0 & -110                 & -87 & -87 & 95 & 89 & -61             & -83 & -15 & 97 & 159 & -74 \\ 
2 & 0 & -90 & 0 & 77 & 0                    & 109 & -85 & -81 & 98 & 95             & 49 & -82 & -8 & 93 & 106 \\
3 & 103 & 0 & -90 & 0 & 77                  & 100 & 106 & -83 & -81 & 100           & 105 & -176 & -78 & -3 & 99 \\
4 & 0 & 103 & 0 & -90 & 0                   & -77 & 105 & 100 & -80 & -83           & -16 & 104 & -171 & -78 & -26 \\
5 & -70 & 0 & 103 & 0 & -90                 & -73 & -85 & 102 & 95 & -82            & -68 & 15 & 106 & -173 & -76 \\
\midrule
$n$ & \multicolumn{15}{c|}{$|Y^{(2)}|$ (mS)} \\
\midrule
1 & 3.46 & 0.00 & 0.50 & 0.00 & 0.13        & 3.30 & 0.05 & 0.39 & 0.06 & 0.10      & 3.26 & 0.01 & 0.41 & 0.05 & 0.10 \\ 
2 & 0.00 & 0.75 & 0.00 & 0.16 & 0.00        & 0.14 & 0.61 & 0.04 & 0.12 & 0.04      & 0.27 & 0.50 & 0.11 & 0.10 & 0.03 \\
3 & 1.49 & 0.00 & 0.22 & 0.00 & 0.08        & 1.26 & 0.05 & 0.16 & 0.03 & 0.02      & 1.24 & 0.01 & 0.17 & 0.02 & 0.03 \\
4 & 0.00 & 0.32 & 0.00 & 0.10 & 0.00        & 0.05 & 0.43 & 0.01 & 0.11 & 0.02      & 0.10 & 0.16 & 0.05 & 0.05 & 0.01 \\
5 & 0.65 & 0.00 & 0.13 & 0.00 & 0.03        & 0.64 & 0.03 & 0.08 & 0.01 & 0.03      & 0.61 & 0.02 & 0.07 & 0.01 & 0.03 \\
\midrule
$n$ & \multicolumn{15}{c|}{$\angle Y^{(2)}$ (Degrees)} \\
\midrule
1 & -77 & 0 & 110 & 0 & -65                 & -85 & -94 & 88 & 89 & -122            & -85 & -72 & 86 & 15 & -104 \\
2 & 0 & 110 & 0 & -65 & 0                   & 140 & 97 & 30 & -98 & -129            & -7 & -75 & -9 & 95 & -69 \\
3 & 110 & 0 & -65 & 0 & 131                   & 103 & -2 & -81 & -144 & 57            & 106 & 14 & -80 & 25 & 61 \\
4 & 0 & -65 & 0 & 131 & 0                   & -58 & -57 & -87 & 137 & 98            & 1 & 117 & -2 & -71 & 99 \\
5 & -65 & 0 & 131 & 0 & -59                 & -71 & 110 & 108 & -38 & 114           & -69 & -10 & 107 & -5 & 116 \\
\bottomrule
\end{tabular*}
\end{table}

The time-periodic Preisach model does not make an appreciable improvement to the accuracy of $Y^{(1)}$ and $Y^{(2)}$ when either $n$ or $m$ are even integers. This is most noticeable in the phase of $Y^{(2)}$ when both $n$ and $m$ are even integers, where the phase is incorrect by as much as 180\textdegree. The exact reason for this inaccuracy is difficult to identify as there is some confidence issues with the perturbation test results. The issue is that the apparent admittances $Y_{n,m,\phi}$ do not closely align with a linear approximation, i.e. the points $Y_{n,m,\phi}$ do not clearly trace a circle in the admittance plane. To quantify the deviation from the ideal circular pattern, the $M_{n,m}$ metric is defined:

\begin{equation}
    M_{n,m} = \frac{1}{N_{\phi} |Y^{(2)}_{n,m}|} \sum_{\phi = 0}^{N_{\phi-1}} \Big(Y_{n,m,\phi} - Y^{(1)}_{n,m} -Y^{(2)}_{n,m}e^{-j4\pi\phi/N_{\phi}}\Big)
\end{equation}

A perfect circular pattern occurs when $M_{n,m}=0$. When $M_{n,m}=1$, the points $Y_{n,m,\phi}$ deviate from the circle by as much as its radius $|Y^{(2)}_{n,m}|$ on average. How close $M_{n,m}$ is to zero determines the level of confidence in the results. The $M_{n,m}$ results for the test transformer in Table \ref{tbl_linear_similarity} shows the odd to odd couplings for $Y^{(1)}$ and $Y^{(2)}$ are most aligned to a linear approximation. The even harmonic to even harmonic couplings do retain a reasonable amount of accuracy, around the $M_{n,m} = 0.5$ range. It appears the time-periodic Preisach model is incomplete in its asymmetric modeling, which is not surprising as it was only constructed from symmetric test data. Future work is recommended to expand data driven curve fitting, especially to improve the common mode function $c(i, \beta_{m}, \alpha_{m})$ to asymmetric operation.

\begin{table}[width=.4\linewidth,cols=6,pos=h]
\caption{$M_{n,m}$ metric for the perturbation results of the tested transformer.}\label{tbl_linear_similarity}
\begin{tabular*}{\tblwidth}{@{} |c|ccccc|@{} }
\toprule
$n$ / $m$ & 1 & 2 & 3 & 4 & 5 \\
\midrule
1 & 0.03 & 1.10 & 0.14 & 1.23 & 0.36 \\
2 & 1.18 & 0.34 & 1.34 & 0.64 & 0.79 \\
3 & 0.07 & 1.50 & 0.22 & 1.26 & 0.68 \\
4 & 1.41 & 0.43 & 1.24 & 0.55 & 0.93 \\
5 & 0.11 & 1.16 & 0.45 & 1.56 & 0.25 \\
\bottomrule
\end{tabular*}
\end{table}

\section{Conclusion}

The harmonic domain linearisation of any nonlinear device is possible if a time-domain model exists, and if the output of the device smoothly changes while varying any harmonic input. Intuitively, this is to be expected. However, the process of linearisation can be difficult to determine. Especially if the nonlinear device is best described by an operator between two sets of functions: the first being the set of possible voltage inputs over time, and the second being for the current output. This work has presented a general linearisation method based on Fourier series definition and the transformation formula for calculating series coefficients. To demonstrate its capability, the methodology was applied to the Preisach model of hysteresis. Improved accuracy in modeling transformer open-circuit characteristics was demonstrated over simpler models of only magnetic saturation. Furthermore, a data-driven methodology for constructing a time-periodic Preisach model was developed. The Preisach model built from symmetric test data provided accurate couplings between odd input and odd output harmonic couplings. However, the absence of asymmetric test data resulted in inaccurate even harmonic couplings.

\section*{Funding sources}

The authors would like to thank the Ministry of Business, Innovation and Employment, New
Zealand. Grant Number: CONT-69662-SSIFAETP-UOC.

\printcredits

\section*{Declaration of competing interest}

Josh Schipper reports financial support was provided by Ministry of Business Innovation and Employment. Co-author has previously been on the editorial board for the journal. -N.W. If there are other authors, they declare that they have no known competing financial interests or personal relationships that could have appeared to influence the work reported in this paper.

\section*{Data availability}

Data collected during the writing of this paper is available upon request from the lead author.

\section*{Acknowledgements}

The authors would like to thank Robert Stiegler and Ana Maria Blanco Castaneda for their support in testing the transformer at Dresden University of Technology. Thank you Gerald Heydt and Jeremy Watson for helpful comments on initial drafts. Thank you Raoul Schipper for proofreading.

\appendix
\section{Generating Function Method}
\label{sec_generating_method}

This appendix presents a general method for constructing the splitting function $d(i, \gamma_{m})$ so that the internal minor loop condition $\mu(\beta, \alpha) \geq 0$ is satisfied. In the process, feasible conditions are developed for  $\lambda_{s}(i, \gamma_{m})$ so that $\mu(\beta, \alpha) \geq 0$. This method is difficult to implement in practice from numerical test data, as it requires differentiation and integration. Therefore, it has not been implemented. 

The first step is to twice differentiate $h(\beta_{m}, \alpha_{m})$ with its definition from (\ref{eq_h_explicit_reduced}):

\begin{equation}
    0 \leq \mu(\beta, \alpha) = \frac{\partial^{2}h}{\partial \beta_{m} \partial \alpha_{m}}(\beta, \alpha) = \frac{1}{2}\left\{ \begin{array}{ll} 
        -\frac{\partial^{2}\lambda_{s}}{\partial i \partial \gamma_{m}}(\alpha, -\beta) + \frac{\partial^{2}d}{\partial i \partial \gamma_{m}}(\alpha, -\beta) & \quad -\beta \geq \alpha \\
        & \\
        \frac{\partial^{2}\lambda_{s}}{\partial i \partial \gamma_{m}}(\beta, \alpha) + \frac{\partial^{2}d}{\partial i \partial \gamma_{m}}(\beta, \alpha) & \quad -\beta \leq \alpha
    \end{array}\right.
\end{equation}

\noindent which simplifies to the expression:

\begin{equation} \label{eq_d_internal_inequality}
    \frac{\partial^{2}d}{\partial i \partial \gamma_{m}}(\beta, \alpha) \geq \left| \frac{\partial^{2}\lambda_{s}}{\partial i \partial \gamma_{m}}(\beta, \alpha) \right|
\end{equation}

The generating function $g(\beta, \alpha) \geq 0$ is added to the right hand side of (\ref{eq_d_internal_inequality}) to make it an equality statement:

\begin{equation} \label{eq_d_internal_equality}
    \frac{\partial^{2}d}{\partial i \partial \gamma_{m}}(\beta, \alpha) = \left| \frac{\partial^{2}\lambda_{s}}{\partial i \partial \gamma_{m}}(\beta, \alpha) \right| + g(\beta, \alpha)
\end{equation}

\begin{figure}
	\centering
		\includegraphics[scale=1]{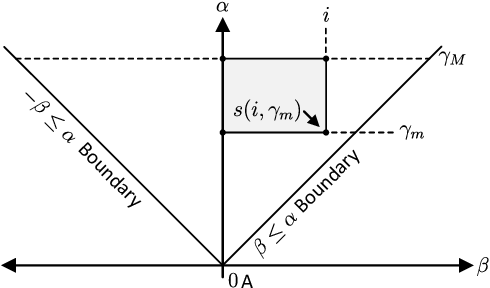}
	\caption{Region of integration to reconstruct $d(i, \gamma_{m})$ in the domain of $\partial^{2}d/\partial i \partial \gamma_{m}$.}
	\label{fig_general_fit}
\end{figure}

The splitting function $d(i, \gamma_{m})$ is reconstructed from (\ref{eq_d_internal_equality}) by integrating over the rectangular region shaded in Fig. \ref{fig_general_fit}:

\begin{equation} \label{eq_d_integrated}
    \int_{\gamma_{m}}^{\gamma_{M}} \int_{0}^{i} \frac{\partial^{2}d}{\partial i \partial \gamma_{m}}(\beta, \alpha) \, d\!\beta \, d\!\alpha = \int_{\gamma_{m}}^{\gamma_{M}} \int_{0}^{i} \left| \frac{\partial^{2}\lambda_{s}}{\partial i \partial \gamma_{m}}(\beta, \alpha) \right| + g(\beta, \alpha) \, d\!\beta \, d\!\alpha = s(i, \gamma_{m})
\end{equation}

\noindent where $\gamma_{M}$ is the maximum value of $\gamma_{m}$ for which the coil is tested, and $s(i, \gamma_{m})$ represents the right hand side integral term.

The left hand side integral of (\ref{eq_d_integrated}) is evaluated and rearranged for $d(i, \gamma_{m})$:

\begin{equation} \label{eq_d_integrated_result}
    d(i, \gamma_{m}) = d_{M}(i) + d_{0}(\gamma_{m}) - d_{0}(\gamma_{M}) - s(i, \gamma_{m})
\end{equation}

\noindent where $d_{M}(i) = d(i, \gamma_{M})$ and $d_{0}(\gamma_{m}) = d(0, \gamma_{m})$. These two functions $d_{M}(i)$ and $d_{0}(\gamma_{m})$ are not determined by (\ref{eq_d_integrated}), but are selected to satisfy some of the conditions in Section \ref{sec_fitting_reduced}. Firstly, it is best to let $d_{0}(\gamma_{m}) = 0$, which also requires $d_{M}(0) = 0$, as $d_{0}(\gamma_{M}) = d_{M}(0)$. The function $d_{M}(i)$ is selected so that $d(-i, i) = 0$ for $i \leq 0$ and $d(i, i) = 0$ for $i \geq 0$, which simplifies (\ref{eq_d_integrated_result}) to:

\begin{equation} \label{eq_d_final}
    d(i, \gamma_{m}) = - s(i, \gamma_{m}) + \left\{\begin{array}{ll} 
    s(i, -i) & \quad i \leq 0 \\
    s(i, i) & \quad i \geq 0 \end{array} \right.
\end{equation}

The next condition to satisfy is (\ref{eq_h_explicit_betam_step4}) by restricting $g(\beta, \alpha)$. Therefore, evaluate (\ref{eq_do_def}) according to (\ref{eq_d_final}):

\begin{equation} \label{eq_do_final}
    d_{o}(i, \gamma_{m}) = - s_{o}(i, \gamma_{m}) + \left\{\begin{array}{ll} 
    s_{o}(i, -i) & \quad i \leq 0 \\
    s_{o}(i, i) & \quad i \geq 0 \end{array} \right.
\end{equation}

\noindent where $s_{o}(i, \gamma_{m}) = (s(i, \gamma_{m}) - s(-i, \gamma_{m}))/2$. For the case $i \geq 0$, $d_{o}(i, \gamma_{m})$ has the following integral form:

\begin{equation}  \label{eq_do_integral_form}
    d_{o}(i, \gamma_{m}) = \frac{1}{2}\int_{i}^{\gamma_{m}} \int_{-i}^{i} \left| \frac{\partial^{2}\lambda_{s}}{\partial i \partial \gamma_{m}}(\beta, \alpha) \right| + g(\beta, \alpha) \, d\!\beta \, d\!\alpha
\end{equation}

Differentiating (\ref{eq_do_integral_form}) with respect to $i$ using the Leibniz integral rule, and evaluating at $i=\gamma_{m}$:

\begin{equation} \label{eq_do_i_diff}
    \frac{\partial d_{o}}{\partial i}(\gamma_{m}, \gamma_{m}) = -\frac{1}{2} \int_{-\gamma_{m}}^{\gamma_{m}} \left| \frac{\partial^{2}\lambda_{s}}{\partial i \partial \gamma_{m}}(\beta, \gamma_{m}) \right| + g(\beta, \gamma_{m})  \, d\!\beta
\end{equation}

Comparing (\ref{eq_h_explicit_betam_step4}) with (\ref{eq_do_i_diff}) it is necessary to restrict $g(\beta, \alpha)$ in the following way:

\begin{equation} \label{eq_g_requirement_1}
    \frac{1}{2} \int_{-\gamma_{m}}^{\gamma_{m}} g(\beta, \gamma_{m})  \, d\!\beta = -\frac{1}{2} \int_{-\gamma_{m}}^{\gamma_{m}} \left| \frac{\partial^{2}\lambda_{s}}{\partial i \partial \gamma_{m}}(\beta, \gamma_{m}) \right| d\!\beta - \frac{\partial \lambda_{se}}{\partial i}(\gamma_{m}, \gamma_{m}) = m_{0}(\gamma_{m})
\end{equation}

For the internal minor loops condition to hold, it is necessary for $g(\beta, \alpha) \geq 0$, and by consequence of (\ref{eq_g_requirement_1}) for $m_{0}(\gamma_{m}) \geq 0$. If $m_{0}(\gamma_{m}) < 0$, then it is impossible for $\mu(\beta, \alpha) \geq 0$. Lastly, without showing the derivation, the second order derivative continuity condition (\ref{eq_d2_continuity_condition}) places the following restriction on $g(\beta, \alpha)$:

\begin{equation} \label{eq_g_requirement_2}
    g_{o}(\gamma_{m}, \gamma_{m}) = \frac{\partial^{2} \lambda_{so}}{\partial i \partial \gamma_{m}}(\gamma_{m}, \gamma_{m}) - \frac{1}{2}\left| \frac{\partial^{2}\lambda_{s}}{\partial i \partial \gamma_{m}}(\gamma_{m}, \gamma_{m}) \right| + \frac{1}{2}\left| \frac{\partial^{2}\lambda_{s}}{\partial i \partial \gamma_{m}}(-\gamma_{m}, \gamma_{m}) \right|
\end{equation}

\noindent where $g_{o}(\beta, \alpha) = (g(\beta, \alpha) - g(-\beta, \alpha))/2$.

\section{Derivation of Approximate Measures}
\label{sec_approximate_measures}

This appendix explains how $M_{j} \geq 0$ for $j = \{1, 2, 3, 4, 5\}$ from (\ref{eq_M_def}) converts to Conditions 1, 4, 5, 6 and 7 of Section \ref{sec_approximate_method}. The conversion of $M_{3} \geq 0$ is given in detail, where (\ref{eq_M_def}) is given integration limits:

\begin{equation}
    M_{3} = \int_{-\gamma_{2}}^{-\gamma_{1}} \int_{\gamma_{1}}^{\gamma_{2}} \mu(\beta, \alpha) \, d\!\alpha \, d\!\beta
\end{equation}

Evaluating the integral with the help of (\ref{eq_general_preisach_weight_formulas}):

\begin{equation}
    M_{3} = h(-\gamma_{2}, \gamma_{1}) - h(-\gamma_{1}, \gamma_{1}) - h(-\gamma_{2}, \gamma_{2}) + h(-\gamma_{1}, \gamma_{2})
\end{equation}

\noindent which can be simplified to $M_{3} = h(-\gamma_{2}, \gamma_{1}) + h(-\gamma_{1}, \gamma_{2})$, because $h(-\gamma, \gamma) = 0$. Evaluate $h$ according to (\ref{eq_h_explicit_reduced}):

\begin{equation}
    M_{3} = \frac{1}{2} \Big( \lambda_{s}(\gamma_{1}, \gamma_{2}) - d(\gamma_{1}, \gamma_{2}) + \lambda_{s}(-\gamma_{1}, \gamma_{2}) + d(-\gamma_{1}, \gamma_{2})\Big)
\end{equation}

Lastly, separate $\lambda_{s}$ and $d$ into odd and even components:

\begin{equation} \label{eq_A6}
    M_{3} = \lambda_{se}(\gamma_{1}, \gamma_{2}) - d_{o}(\gamma_{1}, \gamma_{2}) \geq 0
\end{equation}

\noindent which gives the condition $\lambda_{se}(\gamma_{1}, \gamma_{2}) \geq d_{o}(\gamma_{1}, \gamma_{2})$ for all $\gamma_{2} \geq \gamma_{1} \geq 0$. Applying the same process to $M_{1}$, $M_{2}$, $M_{4}$ and $M_{5}$ gives:

\begin{equation} \label{eq_A4}
    M_{1} = \lambda_{SE}(\gamma_{1}, \gamma_{2}) - \lambda_{so}(\gamma_{1}, \gamma_{2}) - d_{E}(\gamma_{1}, \gamma_{2}) + d_{o}(\gamma_{1}, \gamma_{2}) \geq 0
\end{equation}
\begin{equation} \label{eq_A5}
    M_{2} = -\lambda_{SE}(\gamma_{1}, \gamma_{2}) - \lambda_{so}(\gamma_{1}, \gamma_{2}) + d_{E}(\gamma_{1}, \gamma_{2}) + d_{o}(\gamma_{1}, \gamma_{2}) \geq 0
\end{equation}
\begin{equation} \label{eq_A7}
    M_{4} = -\lambda_{SE}(\gamma_{1}, \gamma_{2}) + \lambda_{so}(\gamma_{1}, \gamma_{2}) - d_{E}(\gamma_{1}, \gamma_{2}) + d_{o}(\gamma_{1}, \gamma_{2}) \geq 0
\end{equation}
\begin{equation} \label{eq_A8}
    M_{5} = \lambda_{SE}(\gamma_{1}, \gamma_{2}) + \lambda_{so}(\gamma_{1}, \gamma_{2}) + d_{E}(\gamma_{1}, \gamma_{2}) + d_{o}(\gamma_{1}, \gamma_{2}) \geq 0
\end{equation}

\noindent where $\lambda_{SE}(\gamma_{1}, \gamma_{2}) = \lambda_{se}(\gamma_{1}, \gamma_{2}) + \lambda_{se}(0, \gamma_{1}) -  \lambda_{se}(0, \gamma_{2})$ and $d_{E}(\gamma_{1}, \gamma_{2}) = d_{e}(\gamma_{1}, \gamma_{2}) + d_{e}(0, \gamma_{1}) - d_{e}(0, \gamma_{2})$. The following implications come from (\ref{eq_A6})-(\ref{eq_A8}):

\begin{equation} \label{eq_A9}
    (M_{1} + M_{2})/2 \geq 0 \text{,  } (M_{4} + M_{5})/2 \geq 0 \text{  and  } M_{3} \geq 0 \text{  implies  } \lambda_{s}(i, \gamma_{m}) \geq 0 \text{  and  } \lambda_{se}(i, \gamma_{m}) \geq |\lambda_{so}(i, \gamma_{m})|
\end{equation}
\begin{equation} \label{eq_A10}
    (M_{1} + M_{2})/2 \geq 0 \text{  and  } (M_{4} + M_{5})/2 \geq 0 \text{  implies  } d_{o}(\gamma_{1}, \gamma_{2}) \geq |\lambda_{so}(\gamma_{1}, \gamma_{2})|
\end{equation}
\begin{equation} \label{eq_A13}
    M_{1} \geq 0 \text{  and  } M_{4} \geq 0 \text{  implies  } d_{o}(\gamma_{1}, \gamma_{2}) - d_{E}(\gamma_{1}, \gamma_{2}) \geq |\lambda_{SE}(\gamma_{1}, \gamma_{2}) - \lambda_{so}(\gamma_{1}, \gamma_{2})|
\end{equation}
\begin{equation} \label{eq_A14}
    M_{2} \geq 0 \text{  and  } M_{5} \geq 0 \text{  implies  } d_{o}(\gamma_{1}, \gamma_{2}) + d_{E}(\gamma_{1}, \gamma_{2}) \geq |\lambda_{SE}(\gamma_{1}, \gamma_{2}) + \lambda_{so}(\gamma_{1}, \gamma_{2})|
\end{equation}

Conditions (\ref{eq_A9}) and (\ref{eq_A10}) along with (\ref{eq_h_explicit_betam_step4}) and (\ref{eq_d2_continuity_condition}) give a good indication that $d(i, \gamma_{m})$ can be derived according to (\ref{eq_d_approx}). The consequences of $m(\zeta)$ being odd are:

\begin{equation} \label{eq_de_def_approx}
    d_{e}(i, \gamma_{m}) = m(i/\gamma_{m})\lambda_{so}(i, \gamma_{m})
\end{equation}
\begin{equation} \label{eq_do_def_approx}
    d_{o}(i, \gamma_{m}) = m(i/\gamma_{m})\lambda_{se}(i, \gamma_{m})
\end{equation}

There are several reasons for selecting $d(i, \gamma_{m})$ according to (\ref{eq_d_approx}). Reason 1, differentiating (\ref{eq_do_def_approx}) with respect $i$ and letting $\hat{m}(1) = 1$ satisfies (\ref{eq_h_explicit_betam_step4}). Note, $\hat{m}(\zeta) \neq 1$ for all $\zeta$ because for $m(\zeta)$ to be an odd function requires $\hat{m}(0) = 0$. Reason 2, twice differentiating (\ref{eq_de_def_approx}) with respect to $i$ and $\gamma_{m}$ and letting $d\hat{m}(1)/d\zeta = 0$ satisfies (\ref{eq_d2_continuity_condition}). Reason 3, from (\ref{eq_A10}) it is noticed that $d_{o}(i, \gamma_{m})$ has to be larger than $|\lambda_{so}(i, \gamma_{m})|$ in the half plane $i \geq 0$. Since magnetic hysteresis loops have a large amount of even symmetry, i.e. $\lambda_{se}(i, \gamma_{m}) \geq |\lambda_{so}(i, \gamma_{m})|$, and because of (\ref{eq_do_def_approx}), $d_{o}(\gamma_{1}, \gamma_{2})$ should have the best potential for satisfying (\ref{eq_A10}).

The remaining conditions upon $\hat{m}(\zeta)$ from Section \ref{sec_approximate_method} are explained:

\begin{itemize}
    \item[4)] (\ref{eq_A6}) requires:
        $$ \lambda_{se}(\gamma_{1}, \gamma_{2}) \geq d_{o}(\gamma_{1}, \gamma_{2}) = \hat{m}(\gamma_{1}/\gamma_{2})\lambda_{se}(\gamma_{1}, \gamma_{2}) $$
        which implies $\hat{m}(\zeta) \leq 1$.
        
    \item[5)] (\ref{eq_A10}) requires $\hat{m}(\gamma_{1}/\gamma_{2})\lambda_{se}(\gamma_{1}, \gamma_{2}) \geq |\lambda_{so}(\gamma_{1}, \gamma_{2})|$, which can be achieved through creating the $m_{1}(\zeta)$ measure.

    \item[6)] (\ref{eq_A13}) requires:

    \begin{equation} \label{eq_m2_temp_1}
    \hat{m}(\gamma_{1}/\gamma_{2})\big(\lambda_{se}(\gamma_{1}, \gamma_{2}) -  \lambda_{so}(\gamma_{1}, \gamma_{2})\big) \geq |\lambda_{SE}(\gamma_{1}, \gamma_{2}) - \lambda_{so}(\gamma_{1}, \gamma_{2})|
    \end{equation}

    To divide (\ref{eq_m2_temp_1}) by $(\lambda_{se}(\gamma_{1}, \gamma_{2}) -  \lambda_{so}(\gamma_{1}, \gamma_{2}))$, this term has to be always positive. This is only possible if $\lambda_{se}(i, \gamma_{m}) \geq |\lambda_{so}(i, \gamma_{m})|$ or equivalently $m_{1}(\zeta) \leq 1$. After the division, it is noticed that $m_{2}(\zeta)$ has the property:

    $$ \hat{m}(\gamma_{1}/\gamma_{2}) \geq m_{2}(\gamma_{1}/\gamma_{2}) \geq \frac{|\lambda_{SE}(\gamma_{1}, \gamma_{2}) - \lambda_{so}(\gamma_{1}, \gamma_{2})|}{\lambda_{se}(\gamma_{1}, \gamma_{2}) -  \lambda_{so}(\gamma_{1}, \gamma_{2})} $$

    \item[7)] A similarly derivation from (\ref{eq_A14}) results in the $m_{3}(\zeta)$ measure.
\end{itemize}

\bibliographystyle{model1-num-names}

\bibliography{cas-refs}





\end{document}